\documentclass[12pt]{article}
\pdfoutput=1
\usepackage{graphicx}
\usepackage{epstopdf}
\usepackage{amsmath}
\usepackage{amsfonts}
\usepackage{amssymb}
\usepackage{color}
\usepackage{mathrsfs}

\usepackage[font={footnotesize}]{caption}

\pdfoutput=1

\usepackage[english]{babel}

\usepackage[colorlinks,bookmarks]{hyperref}
\definecolor{linkblue}{rgb}{0,0,0.8}
\definecolor{linkgreen}{rgb}{0,0.5,0}

\hypersetup{pdfpagemode=UseNone, pdfstartview=FitH, linkcolor=linkblue, %
            citecolor=linkgreen, urlcolor=linkblue}

\numberwithin{equation}{section}

\def\d{{\partial}}

\newcommand{\bea}{\begin{eqnarray}}
\newcommand{\eea}{\end{eqnarray}}
\newcommand{\be}{\begin{equation}}
\newcommand{\ee}{\end{equation}}

\newcommand{\knl}{k_{\rm NL}}
\newcommand{\half}{\frac{1}{2}}

\newcommand{\eqn}[1]{Eq.~(\ref{#1})}

\newcommand{\unitsk}{\, h { \rm Mpc^{-1}}}
\newcommand{\kvec}{\vec{k}}
\newcommand{\xvec}{\vec{x}}
\newcommand{\pvec}{\vec{p}}
\newcommand{\qvec}{\vec{q}}

\newcommand{\momspmeas}[1]{\frac{d^3 \vec{#1}}{(2 \pi)^3} }

\newcommand{\svec}{\vec{s}}

\newcommand{\co}{\mathcal{O}}

\newcommand{\zhat}{\hat{z}}
\newcommand{\khat}{\hat{k}}
\newcommand{\qvecp}{\vec{q'}}

\newcommand{\kdotz}{\hat k \cdot \hat{z}}
\newcommand{\qdotz}{\hat q \cdot \hat z}
\newcommand{\kdotq}{\hat k \cdot \hat q}
\newcommand{\cp}{\mathcal{P}}
\newcommand{\tkvec}{\vec{\tilde{k}}}
\newcommand{\knlv}{  k^r_{ \rm NL}   }

\newcommand{\fnl}{f_{\rm NL}}

\def\xperp{{\vec x}_\perp}
\def\qperp{{\vec q}_\perp}

\def\sperp{{\vec s}_\perp(\vec q,t)}
\def\sperpo{{\vec s}_\perp(\vec 0,t)}

\def\kperp{{\vec k}_\perp}

\def\xparal{{x}_\parallel}
\def\qparal{{q}_\parallel}

\def\sparal{{s}_\parallel(\vec q,t)}
\def\sparalo{{s}_\parallel(\vec 0,t)}
\def\dsparal{{\dot s}_\parallel(\vec q,t)}
\def\dsparalo{{\dot s}_\parallel(\vec 0,t)}
\def\kparal{{k}_\parallel}

\def\epssm{{\epsilon_{s<}}}
\def\epssM{{\epsilon_{s>}}}
\def\epsdm{{\epsilon_{\delta<}}}

\newcommand{\Comment}[1]{{}}

\begin{document}


\setcounter{page}{1} \baselineskip=15.5pt \thispagestyle{empty}

\begin{flushright}
\end{flushright}

\begin{center}

{\Large \bf  On the EFT of Large Scale Structures in Redshift Space}
\\[0.7cm]
{\large Matthew Lewandowski${}^{1,2}$, Leonardo Senatore${}^{1,2}$,\\[0.5cm]
Francisco Prada$^{2,3,4,5}$, Cheng Zhao$^6$ and Chia-Hsun Chuang$^3$}
\\[0.7cm]
\vspace{.3cm}
{\normalsize { \sl $^{1}$ Stanford Institute for Theoretical Physics,\\ Stanford University, Stanford, CA 94306}}\\
\vspace{.3cm}

{\normalsize { \sl $^{2}$ Kavli Institute for Particle Astrophysics and Cosmology, \\
Physics Department and SLAC, Menlo Park, CA 94025}}\\
\vspace{.3cm}

{\normalsize { \sl $^{3}$ Instituto de F{\'i}sica Te{\'o}rica, (UAM/CSIC), \\ Universidad Aut{\'o}noma de Madrid, Cantoblanco,
 E-28049 Madrid, Spain}}\\
\vspace{.3cm}

{\normalsize { \sl $^{4}$ Campus of International Excellence UAM+CSIC, \\
Cantoblanco, E-28049 Madrid, Spain} }\\
\vspace{.3cm}

{\normalsize { \sl $^{5}$ Instituto de Astrof{\'i}sica de Andaluc{\'i}a, (IAA-CSIC), \\ 
Glorieta de la Astronom{\'i}a, E-18190 Granada, Spain}}
\vspace{.3cm}

{\normalsize    {\sl $^{6}$ Tsinghua Center for Astrophysics, Department of Physics, Tsinghua University, Haidian District, Beijing 100084, P. R. China} }

\end{center}

\vspace{.8cm}

\hrule \vspace{0.3cm}
{\small  \noindent \textbf{Abstract} \\[0.3cm]
\noindent  
We further develop the description of redshift space distortions within the Effective Field Theory of Large Scale Structures.  First, we generalize the counterterms to include the effect of baryonic physics and primordial non-Gaussianity. Second, we evaluate the IR-resummation of the dark matter power spectrum in redshift space. This requires us to identify a controlled approximation  that makes the numerical evaluation straightforward and efficient. Third, we compare the predictions of the theory at one loop with the power spectrum from numerical simulations up to $\ell=6$. We find that the IR-resummation allows us to correctly reproduce the BAO peak. The $k$-reach, or equivalently the precision for a given~$k$, depends on additional counterterms that need to be matched to simulations. Since the non-linear scale for the velocity is expected to be longer than the one for the overdensity, we consider a minimal and a non-minimal set of counterterms. The quality of our numerical data makes it hard to firmly establish the performance of the theory at high wavenumbers. Within this limitation, we find that the theory at redshift $z=0.56$ and up to $\ell=2$ matches the data to percent level approximately up to $k \sim 0.13 \unitsk$ or  $k \sim 0.18 \unitsk$, depending on the number of counterterms used, with potentially large improvement over former analytical techniques.  

 \vspace{0.3cm}
\hrule

 \vspace{0.3cm}
 
 \newpage
\tableofcontents

%


\section{Introduction and Summary}

The Effective Field Theory of Large Scale Structures (EFTofLSS)~\cite{Baumann:2010tm,Carrasco:2012cv,Carrasco:2013sva,Carrasco:2013mua,Pajer:2013jj,Carroll:2013oxa,Porto:2013qua,Mercolli:2013bsa,Senatore:2014via,Angulo:2014tfa,Baldauf:2014qfa,Senatore:2014eva,Senatore:2014vja,Lewandowski:2014rca,Mirbabayi:2014zca,Foreman:2015uva,Angulo:2015eqa,McQuinn:2015tva,Assassi:2015jqa,Baldauf:2015tla,Baldauf:2015xfa,Foreman:2015lca,Baldauf:2015aha,Baldauf:2015zga}
 provides the analytical framework that allows us to compute the distribution of dark matter and galaxies at large distances as a perturbative expansion in powers of the overdensity. If successful, this technique has the potential of revolutionizing the way we analyze and extract cosmological information from Large Scale Structure (LSS) data. So far, the EFTofLSS has been compared to simulation data for the case of the dark matter density power spectrum~\cite{Carrasco:2012cv,Carrasco:2013mua,Senatore:2014via,Foreman:2015lca,Baldauf:2015aha} and bispectrum~\cite{Angulo:2014tfa,Baldauf:2014qfa}, the dark matter momentum power spectrum~\cite{Senatore:2014via,Baldauf:2015aha},  the dark matter displacement field~\cite{Baldauf:2014qfa}, the dark-matter vorticity slope~\cite{Carrasco:2013mua,Hahn:2014lca}, the baryon power spectrum~\cite{Carrasco:2013mua}, and the halo power spectrum and bispectra (including all cross correlations with the dark matter field)~\cite{Senatore:2014vja,Angulo:2015eqa}. The results have been very encouraging, showing that the EFTofLSS has a percent level agreement with the numerical data to a much greater wavenumber that formerly available analytic techniques.  

The idea behind the EFTofLSS is that perturbative calculations in LSS are sensitive to short wavelength modes that are not under perturbative control. This induces some `mistakes' in the calculation that need to, and indeed can, be corrected by the insertion of suitable counterterms. This phenomenon happens in several contexts in the EFTofLSS, most notably when describing correlation functions of dark matter in real space, where the counterterms take the form of an effective stress tensor~\cite{Carrasco:2012cv}, for galaxies, where they take the form of bias coefficients~\cite{Senatore:2014eva}, and in redshift space distortions, where they take the form of counterterms for operators involving a product of velocity fields~\cite{Senatore:2014vja}. 

In this paper we provide several developments on the EFTofLSS when applied to dark matter in redshfit space, after reviewing the findings of~\cite{Senatore:2014vja} in Sec.~\ref{review}. In the first part of the paper, in Sec.~\ref{baryonnong}, following the former treatment of non-Gaussian initial conditions in the EFTofLSS~\cite{Angulo:2015eqa,Assassi:2015jqa,Assassi:2015fma}, we describe the general dependence of the redshift space counterterms in the presence of initial non-Gaussianities, extending on the way also the treatment for dark matter and biased tracers. Similarly, following~\cite{Lewandowski:2014rca}, we generalize the counterterms in redshift space to include the effect of baryons and baryonic physics. In both cases, this amounts to allowing counterterms that depend on fields that would not be allowed in the single species case with Gaussian initial conditions, such as the relative velocity between two species or the gravitational potential.

In a second part, in Sec.~\ref{sec:ir-resum}, we perform the IR-resummation of the power spectrum for the various multipoles. This is not quite a trivial task. Though the basic expressions to perform the resummation had been already provided in~\cite{Senatore:2014vja}, their actual numerical implementation would have been very demanding. Instead, we realize that we can perform a {\it controlled} expansion that leads us to expressions where all the integrals can be evaluated in terms of simple analytical formulas. This leads to an expression that is very simple looking. If one is interested in computing the first $\ell$-multipoles of the power spectrum in redshift space at a given $k$, the computational cost is dominated by computing one single FFT for each couple $\{\ell',\ell''\}$ with $0\leq \ell' \leq \ell\,, 0\leq \ell''\leq\ell$. Our expansion is {\it controlled} in the sense that we Taylor expand the exponential of a quantity and evaluated it for the value of the quantity of order one: since the exponential is an analytic function, the Taylor series converges for all values of the quantity. In our case it implies the important fact that  the accuracy of the method can be made arbitrary high by going to higher order in this parameter. In practice we find a very low order is enough. The same trick can be applied also to the IR-resummation of the power spectrum in real space.\footnote{See also~\cite{Baldauf:2015xfa} for a different approximation of the expressions in~\cite{Senatore:2014via} for the IR-resummation in real space.}  Though in that case all the integrals apart for the FFT could be done analytically already with the former methods, the expressions we obtain are much simpler and lead to a much faster evaluation time. We perform the IR-resummation at one-loop order both in real space (to check how well our perturbative scheme works), and in redshift space for the multipoles up to $\ell=6$. The IR-resummation in redshift space works very well: as expected, the residual oscillations in the power spectrum that are present if one does not perform the IR-resummation, are completely eliminated. In particular this implies that the Baryon Acoustic Oscillations (BAO) peak is reproduced to a great accuracy.

In a third part of the paper, in Sec.~\ref{sims}, we move to compare the predictions of the EFTofLSS in redshift space at one-loop against data from numerical simulations. We are using the large volume high-resolution BigMultiDark simulation in the Planck $\Lambda$CDM cosmology at $z = 0 $ and $z = 0.56$, see~\cite{Klypin:2014kpa} for details.\footnote{We look at $z = 0$ because it is when non-linearities are largest, and hence when, for a given $k$, the EFT is most important.  Additionally, we look at $z = 0.56$ because this redshift is more relevant for upcoming surveys, for example BOSS and eBOSS \cite{Dawson:2015wdb}.}  As explained in~\cite{Senatore:2014vja} (see also some earlier partially-related work, as for example~\cite{Vlah:2012ni}), the one-loop perturbative calculation is sensitive to short distance physics through some terms in the equations of motion~(operators). To correct for the mistake associated to assuming that their contribution can be evaluated in perturbation theory, some parameters need to be included in the analytical prediction and need to be measured by matching to the numerical data (renormalization). After reviewing the results of~\cite{Senatore:2014vja}, and after taking the opportunity to  describe some subtleties involving the renormalization of operators involving powers of the velocity  at higher order in Sec.~\ref{review}, we describe how the one-loop prediction depends on four parameters at each redshift: the speed of sounds and its time-derivative, and two additional counterterms that appear directly in redshift space. These parameters can be measured by matching the theory to in principle six power spectra that are predicted at one-loop order: the real space power spectrum and the ones in redshift space up to  $\ell=8$. In practice, the higher multipoles become small and very noisy, making their utility quite limited, so, in practice in this paper we consider only multipoles up to $\ell=6$. In our universe, if we approximate the power spectrum in the $k$-range of interest with a scale free one (an approximation that is just order one correct~\cite{Foreman:2015uva}),  the time-derivative of the speed of sound can be predicted in terms of the speed of sound, reducing the number of free parameters by one. In this approximation, we find that the EFTofLSS in redshift space  matches numerical data up to $\ell=2$ to percent level at $z=0.56$ up to $k \approx 0.13 \unitsk$. When freeing up the time-derivative of the speed of sounds, in Appendix~\ref{freetime}, we find that the $k$-reach is not increased relevantly.  Even though this prediction matches the data sufficiently well,  it does so at the cost of choosing unusually large numerical values of the counterterms. This suggests that the non-linear scale associated to redshift space distortions is longer than for the dark matter in real space. Since the difference between the real space calculation and the redshift space one lies on the renormalization of operators involving the velocity field, this in turn suggests that, at a given redshift, the velocity field becomes non-linear at longer wavelengths than the overdensity field. This is intuitively understandable (see for example~\cite{Tassev:2012cq}): since dark matter moves very slowly in the universe, once the velocity field is non-linear on a given scale, it takes about an Hubble time to move matter around so that the overdensity ceases to be correctly predicted. This fact justifies us to include additional, higher derivative counterterms to the one-loop prediction in redshift space. We explore this in Sec.~\ref{smallkv}, where, with the inclusion of three additional counterterms, we find that the theory matches numerical data up to $\ell=2$ at $z= 0.56$ up to $k \approx 0.18 \unitsk$. 

These results appear to be a remarkable improvement with respect to former analytical techniques (for example SPT at $z= 0.56$ fails at $k \approx 0.05 \unitsk$). Unfortunately, the numerical data that are at our disposal in this work, corresponding to a single realization, appear to be quite noisy and potentially affected by relatively sizable systematic errors, especially at high multipoles. Even though we have actively tried to obtain additional numerical data to better check our findings, the data from the BigMultiDark $(2.5 {\rm Gpc}/h)^3$ simulation box with $3840^3$ particles are the only ones (and therefore the best) we were able to gather. Because of this, performing a percent level matching between theory and data is challenging, and therefore the improvement in the $k$-reach that we find, as well as the numerical values of the counterterms, should be taken with caution.\footnote{Indeed, it would be worthwhile to have additional simulations, including different box sizes, to better establish our results. However, our paper is not just about comparison to simulations, but mainly about the first implementation of the EFT formalism and IR resummation in redshift space and so we leave this to future work. }  In fact, because this study is the first to apply the IR resummation in redshift space and compare with data, we are more interested in confirming that the EFT formalism is working consistently than in determining the precise values of the parameters.  We leave a better determination of the $k$-reach and of the parameters to future work.

Overall, the results of the comparison of the EFTofLSS with data in redshift space are extremely encouraging. It would be interesting to perform higher-order calculations and to compute higher $N$-point functions to better determine the $k$-reach and the accuracy of the theory. Additionally, it would be valuable to compare with different numerical data, potentially from much larger volume, to explore the size and the effect of the numerical systematics.  Finally, it would be worthwhile to compare to other approaches that incorporate non-linear effects in redshift space (see for example \cite{Okumura:2015fga}).

\section{Review of Redshift Space Distortions in the EFTofLSS \label{review}}

We start with a review of the main results of \cite{Senatore:2014vja} (see also references therein, as for example~\cite{Vlah:2012ni}, for some earlier related work). We will work in the ``distant observer'' approximation for simplicity,\footnote{This is reasonable because EFT corrections scale proportional to $k^2 / \knl^2$ (where $\knl$ is the non-linear scale, described below) while full-sky effects, which are related to the curvature of the sphere being observed, scale proportional to $1 /(\chi^2 k^2)$, where $\chi$ is the comoving distance to the measurement.  Furthermore, we can add two additional points. Here we are just comparing to simulations, which do not present us with data as measured on the sphere, but they measure them exactly with the same approximations that we use (flat-sky/distant-observer/plane-parallel). Second, the theory can be improved at high wavenumber without needing to worry about long scale effects (the fact that this is possible is explained by the fact that BOSS data are analyzed in the same approximations as here, and their analysis is limited at short distances by non-linearities, and other effects, and not by the flat-sky approximation).} and we will neglect all general relativistic and evolution effects.  This paper is concerned with the treatment of non-linearities in clustering which are strong on scales much smaller than Hubble, where general relativistic corrections, though easily includable, are negligible. The relation between the real space coordinates $\xvec$ and the redshift space coordinates $\xvec_r$, with line of sight direction $\zhat$, is given by (see for example \cite{Matsubara:2007wj}) 
\be \label{xred}
\xvec_r = \xvec + \frac{\zhat \cdot \vec{v}}{ a H } \zhat \ . 
\ee           
Mass conservation implies that $\rho_r ( \xvec_r ) d^3 \xvec_r = \rho ( \xvec ) d^3 \xvec$, which in Fourier space turns into 
\be 
\delta_r ( \kvec ) = \delta ( \kvec) + \int d^3 x \, e^{- i \kvec \cdot \xvec} \left( \exp \left[ - i \frac{k_z}{aH} v_z ( \xvec) \right] - 1  \right) ( 1 + \delta ( \xvec ) ) \ . 
\ee
In this paper, we will do our computations to one loop, so we can Taylor expand the exponential to cubic order in the fields to obtain
\begin{align} 
\delta_r ( \kvec ) & \simeq \delta ( \kvec ) - i \frac{k_z}{aH} v_z ( \kvec ) + \frac{i^2}{2} \left( \frac{k_z}{aH} \right)^2 [ v_z^2 ]_{\kvec} - \frac{i^3}{3!} \left( \frac{k_z}{aH} \right)^3 [ v_z^3 ]_{\kvec} \nonumber \\
& \hspace{1in} - i \frac{k_z}{aH} [v_z \delta ]_{\kvec} + \frac{i^2}{2} \left( \frac{k_z}{aH} \right)^2 [ v^2_z \delta ]_{\kvec}  \ .  \label{expand}
\end{align}
We have introduced the notation $[ f ]_{\kvec} = \int d^3 x \, e^{- i \vec k \cdot \vec x} f(x)$.  As discussed in \cite{Carrasco:2013mua,Senatore:2014vja}, the velocity field needs to be renormalized because it is really the ratio of two fields, $ v^i ( \xvec , t ) \equiv \pi^i ( \xvec , t ) / \rho(\xvec , t)$.  We do not have control over distances shorter than the non-linear scale, so we cannot take the ratio of two fields at the same point in a controlled way.  It is useful to rewrite the linear piece in terms of the scalar and vector parts of the momentum field, $\pi_S$ and $\pi_V^i$, defined by ($\bar \rho$ is the background energy density)
\be
\pi^i = a \bar \rho \left( \frac{\partial^i }{\partial^2 } \pi_S + \epsilon^{ijk} \frac{\partial_j}{\partial^2} \pi_{V,k} \right)
\ee
where $\partial_i \pi_V^i = 0$.  This is because $\pi^i$ is renormalized directly by $\tau^{ij}$ in the Euler equation of the EFTofLSS, so that, through the continuity equation $\pi_S = - \dot \delta$, $\delta$ is renormalized by the same counterterms as $\pi_S$.  The vector part starts at third order in perturbation theory, and cannot contract with any first order scalar because they behave differently under parity.  Thus, for a one-loop calculation, we do not need any additional counterterms for $\partial_z \pi_z = \bar \rho \partial_z [ ( 1 + \delta ) v_z ] $, and we can use the continuity equation to write \eqn{expand} as 
\begin{align} \nonumber
\delta_r ( \kvec ) & \simeq \delta ( \kvec ) + \frac{k_z^2}{k^2} \frac{ \dot \delta ( \kvec ) }{H} - \epsilon^{z i j} \frac{k_z k_i}{k^2} \frac{\pi_{V,j}}{H} + \frac{i^2}{2} \left( \frac{ k_z}{aH} \right)^2 [v_z^2]_{\kvec}  \\
& \hspace{2in} - \frac{i^3}{3!} \left( \frac{ k_z}{aH} \right)^3 [v_z^3]_{\kvec}  + \frac{i^2}{2} \left( \frac{ k_z}{aH} \right)^2 [v_z^2 \delta]_{\kvec} \ . \label{deltar}
\end{align}
Now we must renormalize the composite operators.  In real space, the equivalence principle implies that the counterterms in the effective stress tensor can only be a function of $\partial_i \partial_j \phi \propto \delta$ and $\partial_i v_j$, and of their spatial derivatives, evaluated on the past trajectory of the fluid element.\footnote{Another covariant object is the momentum equation of motion, $\dot v^i+H v^i+v^j\d_j v^i-\d^i \phi=\rho^{-1}\d_j\tau^{ij}$. But this obviously vanishes on the equations of motions, so it does not contribute to perturbation theory. Of course if one uses the velocity in the counterterm, one should allow for terms to renormalize it. We thank K. Shutz, M. Solon, J. Walsh, and K. Zurek for discussions partially related to this point. }  The latter requirement means that in $\vec x$ space, the operators should be evaluated at the recursively defined fluid position~\cite{Carrasco:2013mua}
\be
\vec x_{\rm fl} ( \xvec , \tau , \tau' ) = \xvec - \int_{\tau'}^\tau d \tau'' \, \vec v ( \vec x_{\rm fl} ( \xvec , \tau , \tau'' ) , \tau'' ) \ .
\ee
Similarly, to renormalize the contact operators in \eqn{deltar}, we can use any operators with the same transformation properties under diffeomorphisms as the bare non-linear term in question, although they should still be evaluated at $\vec x_{\rm fl}$ and integrated along the past history; however, we can expand the operators in the $\vec x_{\rm fl}$ argument to produces a series of higher order operators which are all evaluated at $\xvec$.  Below, we will assume that this has already been done.  

To figure out the general form of the counterterms, first consider the renormalization of the velocity operator
\be
[v^i]_R = v^i + \co^i_v 
\ee
where $\co^i _v$ are the counterterms used.  For $[v^i]_R$ to transform the same way as $v^i$ under a Galilean boost $v^i\rightarrow v^i + \chi^i$, we need $\co^i_v \rightarrow \co^i_v$, i.e. a scalar under Galilean transformations.   Now, we want $[v^i v^j]_R$ to transform the same way as $v^i v^j \rightarrow v^i v^j +  \chi^i v^j+v^i \chi^j + \chi^i \chi^j$.  In order to do this, we must use 
\be
[v^i v^j]_R = v^i v^j +  v^i \co^j_v+ \co_v^i v^j  + \co^{ij}_{v^2} 
\ee
where $\co^{ij}_{v^2}$ are new scalar counterterms.  Thus, we have $[v^iv^j]_R \rightarrow [v^i v^j]_R +  \chi^i [v^j]_R+[v^i]_R  \chi^j+ \chi^i \chi^j$.    In the same way, for cubic terms, we must use, schematically, 
\be
[v^3]_R = v^3 + 3 v^2 \co_v + 3 v \co_{v^2} + \co_{v^3} \ . 
\ee
Because we will only work to one loop in this paper, we will only be adding counterterms that are linear in the fields.  These come from the Galilean-scalar parts $\co^i_v$, $\co^{ij}_{v^2}$, and $\co^{ijk}_{v^3}$.  However, when considering higher order counterterms, one must consider the structure above.  The renormalized operators relevant for this paper, then, are 
\begin{align} \nonumber
[v^2_z]_{R , \kvec} & = \hat{z}^i \hat{z}^j \Bigg\{ [v_i v_j]_{\kvec} + \left( \frac{aH}{\knlv} \right)^2 \left[ c_{11} \delta_{ij} \delta^{(3)}_D ( \kvec) + \left( c_{12} \delta_{ij} + c_{13} \frac{ k_i k_j}{k^2} \right) \delta(\kvec) \right] \Bigg\} \\
& = [v^2_z]_{\kvec} + \left( \frac{aH}{\knlv} \right)^2 \left[ c_{11} \delta^{(3)}_D ( \kvec)  + \left( c_{12}  + c_{13} \mu^2 \right) \delta(\kvec) \right]  \label{renorm1}
\end{align}
\begin{align} \nonumber
[v^3_z]_{R,\kvec} & = \hat{z}^i \hat{z}^j \hat{z}^l \left\{ [v_i v_j v_l ]_{\kvec} + \left( \frac{a H}{\knlv} \right)^2 c_{21} \left( \delta_{ij} v_l + \delta_{i l } v_j + \delta_{j l } v_i  \right)   \right\} \\
& = [v^3_z]_{\kvec} + 3 \left( \frac{ a H}{ \knlv} \right)^2  c_{21}   v_z ( \kvec)  \label{renorm2}
\end{align}
\begin{align} \nonumber
[v^2_z \delta ]_{R , \kvec}   & =    \hat{z}^i \hat{z}^j  \Bigg\{ [v_i v_j \delta]_{\kvec} +  \left( \frac{a H}{\knlv} \right)^2 \left[ c_{31} \delta_{ij} \delta^{(3)}_D ( \kvec) + \left( c_{32} \delta_{ij} + c_{33} \frac{ k_i k_j}{k^2} \right) \delta(\kvec) \right]  \Bigg\}   \\
& = [v^2_z \delta]_{\kvec} +  \left( \frac{a H}{\knlv} \right)^2 \left[ c_{31}  \delta^{(3)}_D ( \kvec)  + \left( c_{32} + c_{33} \mu^2 \right) \delta(\kvec) \right] \label{renorm3} \ 
\end{align}
where $\mu = \kdotz$, and $\delta^{(3)}_D ( \kvec)$ is the Dirac delta function.  In general, the non-linear scale in redshift space, $\knlv$, may be different from the one in real space, $k_{\rm NL}$.  This is because velocities become non-linear before over-densities.  We will discuss this further in Section \ref{smallkv}.  We will also discuss how these expressions are altered in the presence of baryons and primordial non-Gaussianities in Section \ref{baryonnong}.  This analysis was recently done in the study of biased tracers in~\cite{Angulo:2015eqa}, and subsequently extended, for the case of non-Gaussianities, in~\cite{Assassi:2015jqa,Assassi:2015fma}.  The above operators in Eqs.~(\ref{renorm1})-(\ref{renorm3}) should be used in \eqn{deltar} instead of their non-renormalized counterparts, giving
\begin{align} \nonumber
\delta_r ( \kvec) & \simeq \delta( \kvec)  +\mu^2 \frac{\dot{\delta}(\kvec)}{H} - \epsilon^{z i j } \mu \frac{ k_i}{k} \frac{ \pi_V^j }{H}  \\ \nonumber
& \hspace{.2in} + \frac{i^2}{2} \left( \frac{ \mu \, k }{aH} \right)^2 \left(  [v^2_z]_{\kvec} + \left( \frac{aH}{\knlv} \right)^2 \left[ c_{11} \delta^{(3)}_D ( \kvec) + \left( c_{12}  + c_{13} \mu^2 \right) \delta(\kvec) \right]    \right)  \\ \nonumber
& \hspace{.2in} - \frac{i^3}{3!} \left( \frac{ \mu \, k }{aH} \right)^3 \left( [v^3_z]_{\kvec} + 3 \left( \frac{ a H}{ \knlv} \right)^2  c_{21}   v_z ( \kvec)    \right) \\ 
& \hspace{.2in} + \frac{i^2}{2} \left( \frac{ \mu \, k }{aH} \right)^2  \left(  [v^2_z \delta]_{\kvec} +  \left( \frac{a H}{\knlv} \right)^2 \left[ c_{31} \delta^{(3)}_D ( \kvec)  + \left( c_{32} + c_{33} \mu^2 \right) \delta(\kvec) \right]  \right) \ . 
\end{align}
The $c_{11}$ and $c_{31}$ terms renormalize vacuum expectation values, so we can ignore them for the rest of this paper.  We will consider stochastic counterterms, which scale as $(k / \knlv)^4$, in Section \ref{smallkv}.  We can use $v_z ( \kvec) = - i k_z \theta( \kvec) / k^2 $ because vorticity only appears at a higher order in perturbation theory \cite{Carrasco:2012cv,Carrasco:2013mua}, and obtain
\begin{align} \nonumber
\delta_r( \kvec)  & \simeq \delta( \kvec)  +\mu^2 \frac{\dot{\delta}(\kvec)}{H} - \epsilon^{z i j } \mu \frac{ k_i}{k} \frac{ \pi_V^j }{H}  +  \frac{1}{2} \left( \frac{ \mu \, k }{aH} \right)^2  \left( [v^2_z]_{\kvec} + [v^2_z \delta ]_{\kvec} \right)   +  \frac{i}{6} \left( \frac{ \mu \, k }{aH} \right)^3   [v^3_z]_{\kvec} \\
& \hspace{.2in}  - \frac{1}{2} \left( \frac{k}{\knlv} \right)^2 \left(       \left( ( c_{12} + c_{32}) \mu^2 + ( c_{13} + c_{33}) \mu^4 \right) \delta(\kvec)   - \mu^4  c_{21}   \frac{\theta ( \kvec )}{aH}          \right) \ .  \label{deltarr}
\end{align}
Now, letting $f = \partial \log D / \partial \log a$, $2 \pi \bar{c}_1^2 = c_{12} + c_{32}$, and $2 \pi \bar{c}_2^2 = c_{13} + c_{33} + f c_{21}$, and using $\theta_1(\kvec,a) = - f a H \delta_1 ( \kvec , a)$, the power spectrum is 
\bea \label{prred}
&&P^r_{||_\text{1-loop}}(k,\mu,a)= D^2\,P^r_{11}(k,\mu)+D^4\, P^r_{\text{1-loop,\,SPT}}(k,\mu)\\ \nonumber
&&\qquad\qquad -    (2 \pi ) D^2   \Big[     2 c_s^2 +\mu^2\left(4 \, c_s^2\,f +2\frac{d\, c_s^2}{d\log a}\,\right)  +2\, \mu^4    \left( c_s^2\, f^2+\frac{d\, c_s^2}{d\log a}\, f \right)  \\ \nonumber
&&\qquad\qquad\qquad\qquad\qquad\qquad + \left(1+f \mu^2\right)\left(  \bar{c}_1^2 \mu^2+ \bar{c}_2^2 \mu^4\right)   \Big]        \left(\frac{k}{\knl}\right)^2 P_{11}(k) \ ,
\eea
In the EFTofLSS, the time dependence of the counterterm is unknown. However, if the power spectrum was a no-scale power with slope $n$, the additional scaling symmetry would force  $c_s^2$ to have a time dependence $\propto D^{4/(3 + n)}$. In this paper, we use this approximation with $n=-1.5$~\cite{Foreman:2015uva}, to relate $d  c_s^2 / d \log a$ to $c_s^2$ . Given that this approximation is not particular good in our universe, we include a generic time dependence for $c_s$ in Appendix~\ref{freetime}, finding no particularly different results. For explicit expressions for $P_{11}^r$ and $P_{\rm 1-loop, SPT}^r$, see \cite{Matsubara:2007wj}, and for the IR safe versions of the loop integrals, see \cite{Senatore:2014vja} .

\section{Baryons and Primordial non-Gaussianities \label{baryonnong}}

While the previous section has set us up to compare to non-linear data, before we do that, we would like to pause and make a theoretical development.  
Similar to the discussion in \cite{Angulo:2015eqa} in the context of biased tracers, we can straightforwardly account for the effects of baryons and primordial non-Gaussianities when constructing the renormalized operators in Eqs.~(\ref{renorm1})-(\ref{renorm3}).  The results derived in Section \ref{review} depend on three basic assumptions: a single fluid component, general relativity, and Gaussian initial conditions for the fluid.  The assumption of a single fluid meant that the counterterms only depended on dark-matter fields, and the equivalence principle of general relativity was invoked when we said that the counterterms of the effective stress tensor can only depend on $\partial_i \partial_j \phi$, $\partial_i v_j$, and $\vec x_{\rm fl} ( t ) $.  This meant that terms such as $\phi$, $\partial^i \phi$, and $v^i$ were not allowed as counterterms in the effective stress tensor simply because long distance physics cannot affect small scales through these kinds of operators.  The assumption of Gaussian initial conditions was necessary to conclude that long modes affect short scales only through evolution effects, and not through initial correlations.

In this Section, we will extend the analysis in Section \ref{review} to include the presence of baryons and primordial non-Gaussianities by providing the relevant equations.  We will not compare these ideas to simulation data (because they are not available to us), and we will not include the effects of deviations from general relativity; we leave these analyses to later study.

 \subsection{Baryons}
The inclusion of baryonic effects in the EFT of LSS was performed in \cite{Lewandowski:2014rca} where the authors showed that the complex star formation physics happening at short distances can be described, at large distances, by an effective stress tensor and an effective force in the Euler equations for the fluid-like dark-matter and baryonic fields.  The effective force term, which is not present in the case of a single fluid, allows for the possibility of momentum transfer between the two species.  Another new feature is that, along with the alreadty allowable counterterms $\partial^2 \phi \propto w_b \delta_b + w_c \delta_c$, $\partial^i v^j_c$, and $\partial^i v^j_b$ (where $c$ stands for dark matter, $b$ stands for baryon, $w_\sigma = \Omega_\sigma / (\Omega_{c} + \Omega_b )$,  $\sigma = b,c$), the relative velocity $v_b - v_c$ is not forbidden by the equivalence principle in the effective stress tensor and effective force, and so the velocity fields can appear without derivative suppression.  To see why this is the case, consider going to the frame where $\partial \phi = 0$, and let the dark matter and baryon velocities be $v_c$ and $v_b$ respectively in this frame.  Then, we can boost to the center of mass frame, so that in this frame, we have
\begin{align} \nonumber
v_{c,{\rm CM}} & = v_c - (w_c v_c + w_b v_b) = w_b ( v_c - v_b) \\
v_{b,{\rm CM}} & = v_b - (w_c v_c + w_b v_b)  = w_c ( v_b - v_c )
\end{align}
which are both proportional to the relative velocity, and so the counterterms should be able to depend on these fields.  Additionally, notice that for dark-matter correlation functions, the perturbative effects of baryons is expected to be controlled by powers of $w_b \approx .16$ order by order in the expansion.  

In redshift space distortions, the above discussion has two effects.  First, the real space part of the power spectrum is modified through the effective stress tensor as in \cite{Lewandowski:2014rca}.  Second, the renormalized composite operators Eqs.~(\ref{renorm1})-(\ref{renorm3}) will contain extra terms.  Since the number of dark-matter and baryon particles is separately conserved, the expression for the overdensity of each species in redshift space, $\delta_{\sigma, r}$, ($\sigma = b,c$), is the same as \eqn{expand} with all of the bare operators on the right hand side of the same species: 
\begin{align} 
\delta_{\sigma,r} ( \kvec ) & \simeq \delta_\sigma ( \kvec ) - i \frac{k_z}{aH} v_{\sigma,z} ( \kvec ) + \frac{i^2}{2} \left( \frac{k_z}{aH} \right)^2 [ v_{\sigma, z}^{2} ]_{\kvec} - \frac{i^3}{3!} \left( \frac{k_z}{aH} \right)^3 [ v_{\sigma, z}^{3} ]_{\kvec} \nonumber \\
& \hspace{1in} - i \frac{k_z}{aH} [v_{\sigma,z} \delta_\sigma ]_{\kvec} + \frac{i^2}{2} \left( \frac{k_z}{aH} \right)^2 [ v^{ 2}_{\sigma,z} \delta_\sigma ]_{\kvec}  \ .  \label{expand2}
\end{align}
Now, any place where there was a dark-matter field in the renormalization Eqs.~(\ref{renorm1})-(\ref{renorm3}), there can be either a dark-matter or baryon field.  For example, the the new form of the renormalized composite operators for the dark matter fields are
\begin{align}  \nonumber
[v^2_{c,z}]_{R , \kvec} &  = [v^2_{c,z}]_{\kvec}  + w_b c^c_{11} [v_{c,z} v_{b,z}]_{\kvec}  + w_b^2 c^c_{12} [v_{b,z}^2]_{\kvec}  \\
& \hspace{.5in} + \left( \frac{aH}{\knlv} \right)^2  \left[ \left( c^c_{13}  + c^c_{14} \mu^2 \right) \delta_c(\kvec)   +    w_b \left( c^c_{15}  + c^c_{16} \mu^2 \right) \delta_b(\kvec)  \right]\label{renorm12} \\
[v^3_{c,z}]_{R,\kvec} & = [v^3_{c,z}]_{\kvec} + c^c_{21}w_b [v^2_{c,z} v_{b,z}]_{\kvec}  + c^c_{22} w_b^2[v_{c,z} v^2_{b,z}]_{\kvec}  + c^c_{23}w_b^3[ v^3_{b,z}]_{\kvec}   \nonumber \\
& \hspace{.5in} + 3 \left( \frac{ a H}{ \knlv} \right)^2 \left[   c^c_{24}   v_{c,z} ( \kvec)  + w_b c^c_{25}   v_{b,z} ( \kvec)  \right] \label{renorm22} \\
[v^2_{c,z} \delta_c ]_{R , \kvec} & = [v^2_{c,z} \delta_c]_{\kvec} + c^c_{31}w_b [v^2_{c,z} \delta_b]_{\kvec} + c^c_{32}w_b [v_{c,z} v_{b,z} \delta_c]_{\kvec} +  c^c_{33}w_b^2 [v^2_{c,z}  \delta_c]_{\kvec} + c^c_{34}w_b^3 [v^2_{b,z}  \delta_b]_{\kvec}  \nonumber  \\
& \hspace{.5in} +  \left( \frac{a H}{\knlv} \right)^2 \left[ \left( c^c_{35} + c^c_{36} \mu^2 \right) \delta_c(\kvec)  + w_b \left( c^c_{37} + c^c_{38} \mu^2 \right) \delta_b(\kvec)\right] \label{renorm32} \ .
\end{align}
The factors of $w_b$ have been chosen above so that there is a sensible expression in the $w_b \rightarrow 0$ limit. The equations for the baryon operators can be obtained by swapping the $c$ and $b$ labels.

\subsection{Primordial non-Gaussianities}
Everywhere else in this paper, we have assumed that the initial conditions coming from the primordial curvature perturbation were Gaussian, so that all modes were uncorrelated when they entered the horizon.  When this is the case, and since gravitational evolution follows from local interactions, all of the correlation functions must have a $k$ dependence which is consistent with this.  Thus, the $k$ dependence of the counter terms was restricted, for example by not including any $\partial^i/\partial^2$ type terms.  However, when the initial conditions are non-Gaussian, modes of different sizes are correlated when they enter the horizon, and so a different kind of $k$ dependence is allowed by this type of non-locality.  Effects of this type have been discussed in the context of the EFTofLSS in \cite{Angulo:2015eqa,Assassi:2015jqa,Assassi:2015fma}.  

The EFT of large scale structures relies on a hierarchy of scales $k_L / k_S \ll 1$, where $k_S \sim \knl$ is the non-linear scale of structure formation.  This hierarchy allows the effects of short scales on large scales to be described by an expansion in $k_L / k_S$.   Of particular interest when integrating out the short scales, then, is how non-Gaussianities in the short modes will affect the long modes.  Thus, this operation is sensitive to the squeezed limit $k_L / k_S \ll 1$ of various correlation functions of the non-Gaussian field.  In order to discover new possible counterterms, we need to know how averaging over short modes in the background of long modes can effect the long modes.  Because of the non-Gaussian initial conditions, the short wavelength modes depend on the long modes, even before non-linear gravitational evolution is taken into account. We can therefore write them in the following form~(see for example~\cite{Angulo:2015eqa, Assassi:2015jqa,Assassi:2015fma}):  
\be \label{zetang}
\zeta^s_{NG} ( \xvec ) \simeq \zeta^s_g ( \xvec ) + \fnl \int_{\kvec} \int_{\pvec} W ( \kvec , \pvec ) \, \zeta^s_g ( \pvec)\,  \zeta^l_g ( \kvec) \, e^{ i \xvec \cdot ( \kvec + \pvec) } 
\ee
where $\zeta$ is the curvature perturbation, which is constant outside of the horizon.\footnote{The proof that $\zeta$ is constant at the quantum level can be found in \cite{Pimentel:2012tw,Senatore:2012ya}.}  The subscript $g$ means that the field has Gaussian statistics, and the superscripts $s$ and $l$ denote fields with support only at short or only at long wavelengths, respectively.  The function $W( \kvec , \pvec)$ can only depend on $k$, $p$, and $\khat \cdot \hat p$ due to statistical isotropy and homogeneity.  For the purposes of this discussion, we will focus on the case where the largest effect of the non-Gaussianity is on the three-point function.  In this case, in the squeezed limit $k / p \ll 1$, the kernel $W$ can be expanded like 
\be \label{wfunct}
W( \kvec , \pvec ) = \sum_{\ell, i} w_{\ell , i} \left( \frac{k}{p} \right)^{\Delta_{\ell , i}} \cp_{\ell} ( \khat \cdot \hat p)
 \ee
where $w_{\ell , i}$ and $\Delta_{\ell , i}$ are constants, $\cp_\ell$ are the Legendre polynomials, and we will use the $i=0$ index to denote the smallest $\Delta_{\ell , i}$ for any given $\ell$, i.e. to denote the dominant scaling for $k/ p \ll 1$.  The powers $\Delta_{\ell , i}$ depend on the particular inflationary model: $\Delta_{0,0} = 0 $ for local type non-Gaussanity; $\Delta_{0,0} = 2$ for equilateral \cite{Creminelli:2005hu} and orthogonal type \cite{Senatore:2009gt}; $0 \leq \Delta_{0,0} \leq 3/2$ for models of quasi-single field inflation \cite{Chen:2009we,Chen:2009zp}; and $0 \leq \Delta_{0,0} \leq 2$ in strongly coupled conformal theories \cite{Green:2013rd}. Fields with non-zero spin $\ell > 0$ during inflation can give rise to angular dependence in the squeezed limit such that $w_{\ell >0 , i } \neq 0$ \cite{Arkani-Hamed:2015bza}.  Notice that, contrary to what formerly stated in the literature in the context of the dark matter stress tensor and biases, also odd multipoles of $\ell$ are expected to contribute.\footnote{  Odd values of $\ell$ can contribute not only in redshift space, but also in the bias and dark matter stress-tensor expansions.  To see this, consider the expression for the number density of collapsed objects $n_h ( \xvec , t)$, which is a function of the correlation functions of the short modes evaluated on the past path~\cite{Senatore:2014eva}
\be\label{eq:biased}
n_h ( \xvec , t) = \int^t d t' \, K( t , t' ) \, f_h \left(  \langle \delta_s^2 \rangle_l ( \xvec_{\rm fl} ( t , t' ) , t' ) , \langle \delta_s \partial_i v^i_s \rangle_l ( \xvec_{\rm fl} ( t , t' ) , t' ) , \langle \delta_s^3 \rangle_l ( \xvec_{\rm fl} ( t , t' ) , t' ) , \dots            \right)
\ee
where the brackets $\langle \rangle_l$ mean that we average over the short modes in the background of the long modes.  Then, in order for an odd multipole of $W( \kvec , \qvec)$ to contribute in an expression like $\langle \delta_s^2 (\xvec) \rangle_l$, it needs to be multiplied by some other function which is odd under $\kvec \rightarrow - \kvec$ and $\pvec \rightarrow - \pvec$, otherwise it would integrate to zero.  For an effect linear in $\fnl$, this means that it must mix with a short scale gravitational non-linearity, for instance with the $F_2$ and $F_3$ kernels used in perturbation theory.  Generally, the odd parts of $W$ can multiply odd parts in $F_2$ or $F_3$ and contribute to the bias expansion.  

The odd multipoles of $W$ cannot be extracted from the primordial bispectrum because it only depends on the even part in the squeezed limit $k / k_1 \ll 1$
\be
B( k_1 , k_2 , k )\, \rightarrow \, 2 \fnl  \left( W(\kvec_1 , \kvec) + W( -\kvec_1 , \kvec ) \right) P(k_1) P(k) \ , 
\ee
However, the odd multipoles do contribute to the four-point function, as for example from this expression where we include the $f_{\rm NL}$ contribution twice:
\be\label{eq:4pointfnlsq}
\langle \zeta^s ( \kvec_1 ) \zeta^s ( \kvec_2 ) \zeta^l ( \qvec_1 ) \zeta^l ( \qvec_2 ) \rangle \supset \fnl^2 W( \kvec_1 + \qvec_1 , - \qvec_1) W( \kvec_2 + \qvec_2 , - \qvec_2 ) P( k_1 ) P(q_1 ) P(q_2) \ . 
\ee
Therefore, in order to extract the odd pieces of $W$ from a model of inflation, one can compute the four-point function that arises from the insertion of two three-point vertices, matching in this way the $\fnl^2$ part in (\ref{eq:4pointfnlsq}) from \eqn{zetang}. For example these odd spin pieces can arise from the exchange of a spinning particle during inflation~\cite{Arkani-Hamed:2015bza}.  

}

To find the new types of counterterms allowed in redshift space, we plug \eqn{wfunct} into the last term of \eqn{zetang} and write the Legendre polynomials like 
\be
\cp_\ell ( x ) = \sum_{\tilde \ell = 0}^\ell \rho_{\ell , \tilde \ell} \,  x^{\tilde \ell}
\ee
to get 
\begin{align} \label{manips1}
& \int_{\kvec} \int_{\pvec} W ( \kvec , \pvec ) \, \zeta^s_g ( \pvec)\,  \zeta^l_g ( \kvec) \, e^{ i \xvec \cdot ( \kvec + \pvec) }  \\   \nonumber
 & \hspace{.3in}  =   \sum_{\ell, j}  \int_{\kvec} \int_{\pvec} w_{\ell , j} \left( \frac{k}{p} \right)^{\Delta_{\ell , j}} \cp_{\ell} ( \khat \cdot \hat p)   \, \zeta^s_g ( \pvec)\,  \zeta^l_g ( \kvec) \, e^{ i \xvec \cdot ( \kvec + \pvec) }  \\  \nonumber
 & \hspace{.3in} =  \sum_{\ell, j}  \int_{\kvec} \int_{\pvec} w_{\ell , j } \left( \frac{k}{p} \right)^{\Delta_{\ell , j}} \sum_{\tilde \ell = 0}^\ell \rho_{\ell , \tilde \ell} \,  ( \khat \cdot \hat p)^{\tilde \ell}  \, \zeta^s_g ( \pvec)\,  \zeta^l_g ( \kvec) \, e^{ i \xvec \cdot ( \kvec + \pvec) }  \\   \nonumber
 & \hspace{.3in} =  \sum_{\ell, j}  w_{\ell , j }   \int_{\kvec}\left( \frac{k}{\knl} \right)^{\Delta_{\ell , j}}   e^{ i \xvec \cdot \kvec} \zeta^l_g ( \kvec)  \left[ \sum_{\tilde \ell = 0}^\ell \rho_{\ell , \tilde \ell} \,   \khat^{i_1} \cdots \khat^{i_{\tilde \ell}} \right]    \int_{\pvec}  \hat{p}^{i_1}\cdots \hat{p}^{i_{\tilde \ell}}  \left( \frac{\knl}{p} \right)^{\Delta_{\ell , j}}  \zeta^s_g ( \pvec) \, e^{ i \xvec \cdot  \pvec }    
\end{align}
where the sum over repeated $i_1, \dots, i_n$ is implied throughout the rest of this section.  Then, by multiplying by $1 = \delta^{ij} \hat p^i \hat p^j$ the appropriate number of times, we can write 
\begin{align}
\left[ \sum_{\tilde \ell = 0}^\ell \rho_{\ell , \tilde \ell} \,   \khat^{i_1} \cdots \khat^{i_{\tilde \ell}} \right]  \hat{p}^{i_1}\cdots \hat{p}^{i_{\tilde \ell}} = \left[ \sum_{\tilde \ell = 0}^\ell \rho_{\ell , \tilde \ell} \,   \khat^{i_1} \cdots \khat^{i_{\tilde \ell}}\,   \delta^{i_{\tilde \ell} + 1, i_{\tilde \ell} + 2} \cdots  \delta^{i_{\ell} - 1, i_{\ell}}  \right]  \hat{p}^{i_1}\cdots \hat{p}^{i_{ \ell}} 
\end{align}
for any unit vector $\hat p$ (so that the right hand side has an $\ell$-number of $\hat p$'s), and finally define the the symmetric projection operator\footnote{Here, we use the standard notation
\be
V^{( i_1 \cdots i_N)} = \frac{1}{N ! } \sum_{\substack{\text{all perms. of}\\ i_1 , \dots , i_N} } V^{i_1 \cdots i_N} \ . 
\ee}
\be
\sigma(\khat)^{i_1 \cdots i_\ell}_\ell =  \sum_{\tilde \ell = 0}^\ell \rho_{\ell , \tilde \ell} \,  \,  \khat^{( i_1} \cdots \khat^{i_{\tilde \ell}}\,   \delta^{i_{\tilde \ell} + 1, i_{\tilde \ell} + 2} \cdots  \delta^{i_{\ell} - 1, i_{\ell} )} \ . 
\ee
Thus, we have been led to write the Legendre polynomials as
\be 
\cp_\ell ( \khat \cdot \hat{p} ) = \sigma(\khat)^{i_1 \cdots i_\ell}_\ell  \hat{p}^{i_1} \cdots \hat{p}^{i_\ell} \ . 
\ee
The projection operators are easy to write down using the Legendre polynomial coefficients $\rho_{\ell , \tilde \ell}$.  For example, with two of the Legendre polynomials being $\cp_2 ( \mu ) = (3 \mu^2 - 1) / 2$, and $\cp_4 ( \mu )=( 35 \mu^4-30\mu^2+3) / 8$, the two corresponding projection operators are 
\begin{align}
\sigma(\khat)_2^{i j } & = \half \left(  3 \khat^i \khat^j - \delta^{ij} \right)\\
 \sigma(\khat)_4^{ijkl}&  = \frac{1}{8} \left( 35 \,  \khat^i \khat^j \khat^k \khat^l - 30 \frac{1}{6} \left( \khat^i \khat^j \delta^{k l } + \text{5 perms.} \right) + 3 \frac{1}{3} \left( \delta^{ij} \delta^{kl} + \text{2 perms.} \right) \right) \ . 
\end{align}
This definition of $\sigma ( \khat)^{i_1 \cdots i_\ell}_\ell$ ensures that the operators are symmetric under the exchange of any of the indices, and the trace over any two indices is zero.  It also satisfies $\sigma ( \khat)^{i_1 \cdots i_\ell}_\ell \khat^{i_{\ell}} = \sigma( \khat)^{i_1 \cdots i_{\ell - 1}}_{\ell - 1}$.  With these definitions, we can go back to \eqn{manips1} and write 
\begin{align} \nonumber
 & \int_{\kvec} \int_{\pvec} W ( \kvec , \pvec ) \, \zeta^s_g ( \pvec)\,  \zeta^l_g ( \kvec) \, e^{ i \xvec \cdot ( \kvec + \pvec) }  \\ \nonumber
  & \hspace{.3in} =  \sum_{\ell, j}  w_{\ell , j }   \int_{\kvec}   e^{ i \xvec \cdot \kvec}  \left( \frac{k}{\knl} \right)^{\Delta_{\ell , j}} \zeta^l_g ( \kvec) \, \sigma(\khat)_\ell^{i_1 \cdots i_\ell}    \int_{\pvec}  \hat{p}^{i_1}\cdots \hat{p}^{i_{\ell}} \, e^{ i \xvec \cdot  \pvec }\,  \left( \frac{\knl}{p} \right)^{\Delta_{\ell , j}}  \zeta^s_g ( \pvec)    \\ 
  &  \hspace{.3in} =  \sum_{\ell, j}  w_{\ell , j } \,  \psi ( \xvec )_{\ell, j}^{i_1 \cdots i_\ell } \cdot \alpha(\xvec )^{i_1 \cdots i_\ell}_{\ell, j} 
\end{align}
where
\begin{align}
 \psi ( \xvec )_{\ell, j}^{i_1 \cdots i_\ell } & =   \int_{\kvec}  e^{ i \xvec \cdot \kvec}  \left( \frac{k}{\knl} \right)^{\Delta_{\ell , j}} \zeta^l_g ( \kvec) \sigma(\khat)_\ell^{i_1 \cdots i_\ell}      \\
  \alpha(\xvec )^{i_1 \cdots i_\ell}_{\ell, j}  & =  \int_{\pvec} e^{ i \xvec \cdot  \pvec }   \left( \frac{\knl}{p} \right)^{\Delta_{\ell , j}}  \zeta^s_g ( \pvec) \,   \hat{p}^{i_1}\cdots \hat{p}^{i_{\ell}} 
\end{align}
and their Fourier transforms are 
\begin{align}
  \psi ( \kvec )_{\ell, j}^{i_1 \cdots i_\ell } & =  \left( \frac{k}{\knl} \right)^{\Delta_{\ell , j}} \zeta^l_g ( \kvec) \, \sigma(\khat)_\ell^{i_1 \cdots i_\ell}      \\
  \alpha(\pvec )^{i_1 \cdots i_\ell}_{\ell, j}  & =   \left( \frac{\knl}{p} \right)^{\Delta_{\ell , j}}  \zeta^s_g ( \pvec) \,   \hat{p}^{i_1}\cdots \hat{p}^{i_{\ell}} \ . 
\end{align}
Notice that $\alpha$ contains all of the short modes, and $\psi$ contains the long modes. The final expression, in real space, is 
\be \label{zetang1}
\zeta^s_{NG} ( \xvec ) \simeq \zeta^s_g ( \xvec ) + \fnl  \sum_{\ell, j}  w_{\ell , j } \,  \psi ( \xvec )_{\ell, j}^{i_1 \cdots i_\ell }  \alpha(\xvec )^{i_1 \cdots i_\ell}_{\ell, j} 
\ee
which in Fourier space turns into 
\be \label{zetang1four}
\zeta^s_{NG} ( \kvec_S ) \simeq \zeta^s_g ( \kvec_S ) + \fnl \sum_{\ell, j}  w_{\ell , j } \int_{\kvec'_L \ll \kvec_S} \,  \psi ( \kvec'_L )_{\ell, j}^{i_1 \cdots i_\ell }  \alpha(\kvec_S - \kvec'_L )^{i_1 \cdots i_\ell}_{\ell, j} \ .
\ee

When the mode of interest re-enters the horizon, even before it undergoes non-linear gravitational evolution, it will already have non-Gaussian statistics
\be
\delta^{(0)}_{NG} ( \kvec_S , t_{\rm in} ) \simeq \delta_g ( \kvec_S , t_{\rm in}) + \fnl    \sum_{\ell, j}  w_{\ell , j }  \int_{\kvec'_L \ll \kvec_S} \, \tilde \phi ( \kvec'_L , t_{\rm in} )^{i_1 \cdots i_\ell}_{\ell , j}   \, \tilde \delta_g (\kvec_S - \kvec'_L , t_{\rm in})^{i_1 \cdots i_\ell}_{\ell, j}  \ , 
\ee
where 
\be
\tilde \phi( \kvec_L , t_{\rm in} )^{i_1 \cdots i_\ell}_{\ell , j}  = \sum_{i} \frac{5}{2} \frac{H_0^2 \Omega_m}{D(a_{\rm in} ) } \frac{1 }{k_L^2 T(k_L , t_{\rm in}) } \left( \frac{k_L}{\knl} \right)^{\Delta_{\ell,j}} \sigma(\khat_L)_\ell^{i_1 \cdots i_\ell} \, \delta_g ( \kvec_L , t_{\rm in} ) 
\ee
and
\be
\tilde \delta_g( \kvec_S, t_{\rm in} )^{i_1 \cdots i_\ell}_{\ell , j} =\frac{2}{5} \frac{k_S^2}{\Omega_m H_0^2} D(a_{\rm in}) T(k_S , t_{\rm in}) \alpha( \kvec_S )^{i_1 \cdots i_\ell}_{\ell , j} \ . 
\ee
Here, $T(k, t_{\rm in})$ is the transfer function which linearly evolves modes from horizon re-entry until the time $t_{\rm in}$, and $\delta^{(0)}_{NG}$ should now be used as the initial condition for the overdensity field. Thus, when we average over short modes to define the effective field theory, correlation functions will depend on $ \tilde \phi( \xvec )_{\ell, j}^{i_1 \cdots i_\ell } $.  In particular, we should include this field, potentially contracted with other fields to form operators with the desired transformation properties, in the counterterms for the dark matter stress tensor and in the bias terms,~as in~\cite{Angulo:2015eqa, Assassi:2015jqa,Assassi:2015fma} (and, as we point out, with the inclusion of odd multipoles). In particular, these fields a as evaluated at the initial time $t_{\rm in}$ and at the position corresponding to the fluid element at time $t'$ of the past trajectory, $ \tilde \phi( \xvec_{\rm fl}(t,'t_{\rm in}), t_{\rm in })_{\ell, j}^{i_1 \cdots i_\ell }$, while the operators induced by the non-linear evolution appear as evaluated on all times of the past trajectory of the fluid element, $\delta(\xvec_{\rm fl}(t',t_{\rm in}), t' )$. In redshift space, the linear counterterms at lowest order in derivatives enter in the same form (but with different possible coupling constants $c_i^n$) for $[v^2_z]_{R , \kvec} $ and $[v^2_z \delta ]_{R , \kvec} $: 
\begin{align} 
[v^2_z]_{R , \kvec}\text{ , } [v^2_z \delta ]_{R , \kvec}  & \supset \hat{z}_i \hat{z}_j  \left( \frac{aH}{\knl} \right)^2  \sum_n \left[   c_{1}^n \delta^{ij} \tilde \phi ( \kvec)_{0,n}  + c^n_{2} \tilde \phi( \kvec )^{ij}_{2,n}  \right]  \\ \nonumber
& = \left( \frac{aH}{\knl} \right)^2 \frac{5}{2} \frac{H_0^2 \Omega_m}{D(a_{in} ) } \frac{1 }{k^2 T(k , t_{in}) }  \\
& \hspace{.5in} \times  \sum_n \left(  c_1^n  \left( \frac{k}{\knl} \right)^{\Delta_{0 ,n}}   +  c_2^n  \left( \frac{k}{\knl} \right)^{\Delta_{2 ,n}} \cp_2 ( \mu_k )      \right) \delta_g ( \kvec , t_{\rm in} )  \nonumber 
\end{align}
\begin{align}
[v^3_z]_{R , \kvec}  & \supset    \hat{z}_i \hat{z}_j \hat{z}_l  \left( \frac{a H}{\knlv} \right)^3 \sum_n c^n_{3}\,  \tilde \phi(\kvec)^{ijl}_{3,n} \\ \nonumber
& =   \left( \frac{ a H}{ \knlv} \right)^3  \frac{5}{2} \frac{H_0^2 \Omega_m}{D(a_{in} ) } \frac{1 }{k^2 T(k , t_{in}) }  \sum_n c^n_3   \left( \frac{k}{\knl} \right)^{\Delta_{3,n}} \cp_3 ( \mu_k ) \,   \delta_g ( \kvec , t_{\rm in} ) \ .  
\end{align} 
At higher derivative or higher field powers, terms like $k^i \delta^{jk} \tilde \phi ( \kvec)_{0,n}$ and $v^i  \delta^{jk} \tilde \phi ( \kvec)_{0,n}$ would contribute.  No multipoles higher than $\ell = 3$ can contribute at this order because contracting with $k^i$ or $\partial^i \delta$ makes it higher order, and contracting its own indices gives zero because all of the operators are traceless.   Notice that, contrary to non-Gaussianities in dark matter and bias power spectra \cite{Assassi:2015jqa,Assassi:2015fma}, the $\ell=2$ multipole has a distinct signature linear with $\delta$ in redshift space.  The reason that this does not happen in the other cases is the following.  In the stress tensor for dark matter, the linear terms that we can write are 
\be \label{tauijng}
\langle \tau^{ij}  \rangle \supset c_1^n \delta^{ij} \tilde \phi ( \kvec )_{0,n} + c_2^n \tilde \phi ( \kvec )^{ij}_{2,n}
\ee
but what actually enters the Euler equation is $k_i k_j \langle \tau^{ij}  \rangle$, and because 
\be
k_i k_j \tilde \phi ( \kvec )^{ij}_{2,n} = \frac{2}{3} k^2 \left( \frac{k}{\knl} \right)^{\Delta_{2,n} - \Delta_{0,n}} \tilde \phi ( \kvec )_{0,n}
\ee
this term is a higher derivative of the monopole.  Thus, the two terms in \eqn{tauijng} combine to give

\be
k_i k_j \langle \tau^{ij}  \rangle \supset   k^2 \left(  c_1^n  + \frac{2}{3}  c_2^n\left( \frac{k}{\knl} \right)^{\Delta_{2,n} - \Delta_{0,n}}  \right)   \tilde \phi ( \kvec )_{0,n}
\ee
which, under most circumstances, would be difficult to differentiate from a pure scaling. Similarly, odd multipoles usually appear multiplied by the derivative of another operator, in the form for example given by $ \tilde \phi( \xvec_{\rm fl}(t',t_{\rm in}), t_{\rm in } )_{1, j}^{i } \d_i \delta(\xvec_{\rm fl}(t',t_{\rm in}), t') /k_{\rm NL}$.

\section{IR-resummed Power Spectrum in Redshift Space\label{sec:ir-resum}}

The beginning of this section will closely follow previous work \cite{ Senatore:2014vja, Senatore:2014via, {Senatore:2014eva}}, although we have a slightly different presentation tailored for the subsequent discussion.  We start in the Lagrangian picture, where the overdensity in real space is given by 
\be
1 + \delta(\vec x,t)=\int d^3q\; \delta^{(3)}(\vec x-\vec \Psi (\vec q,t))=\int d^3q\;\int \frac{d^3k}{(2\pi)^3}\; e^{ i\,\vec k\cdot\left(  \xvec - \vec q  -  \vec s(\vec q,t)\right)}\ ,
\ee
and the displacement field $\svec$ is defined through the Lagrangian coordinate $\vec{\Psi}$ by $ \vec{\Psi} ( \qvec , t ) = \qvec + \svec ( \qvec , t) $.  In redshift space, using $\vec v = a \dot{\vec \Psi}$, \eqn{xred} implies that $\vec \Psi^r = \vec \Psi + \zhat ( \zhat \cdot \dot{\vec \Psi}/H )$, so the above formula generalizes to 
\bea
&&1 + \delta_r(\vec x_r,t)=\int d^3q\; \delta^{(2)}\left(\xperp-   \vec{\Psi}_\perp  \right) \;\delta^{(1)}\left( \xparal- \Psi_\parallel  -\frac{\dot{ \Psi}_\parallel }{H}\right)\\ \nonumber
&&\qquad\qquad=\int d^3 q   \;\int \frac{d^2k_\perp}{(2\pi)^2} \frac{dk_\parallel}{(2\pi)}\; e^{ i\,\kperp\cdot\left( \vec{x}_{\perp} - \qperp  -  \sperp\right)}\;e^{ i\,\kparal\cdot\left(  x_{||} - \qparal -  \sparal -\frac{\dsparal}{H}\right)}\ .
\eea
We will use the notation $V_z$ and $V_\parallel$ interchangeably to mean $\vec{V} \cdot \zhat$.  Fourier transforming gives, for $\kvec \neq 0$,
\be
\delta_r(\kperp,\kparal,t)=\int d^3 q     \; e^{- i\,\kperp\cdot\left(\qperp+\sperp\right)}\;e^{- i\,\kparal\cdot\left(\qparal+\sparal+\frac{\dsparal}{H}\right)}\ .
\ee
To compute the power spectrum, we take the expectation value $\langle\delta_r(\kperp,\kparal,t)\delta_r(\kperp',\kparal',t')\rangle$ and perform the integral with respect to the average Lagrangian coordinates. Translation invariance in $q$-space ensures an overall factor of $\delta^{(2)}(\kperp+\kperp')\delta^{(1)}(\kparal+\kparal')$ after this integration.  All of this allows us to write the power spectrum as
\bea
&&P^r(k_\perp,\kparal,t) =\int d^3 q   \;  e^{-i \kvec \cdot \qvec}    \left\langle e^{-i\, \kperp\cdot \left(\sperp-\sperpo\right)-i\, \kparal\cdot\left(\sparal-\sparalo+\frac{\dsparal-\dsparalo}{H}\right)}\right\rangle\ .
\eea
By using the cumulant theorem, we can write
\be \label{cumulant}
P^{r}(k_\perp,\kparal,t)=\int d^3q \;  e^{-i \kvec \cdot \qvec} \; K_r(\kvec , \qvec ;t)\ ,
\ee
with
\bea
&&\!\!\!\!\!\!\!\!\!\!\!\! K_r(\kvec , \qvec ;t)=\\ \nonumber
&&\!\!\!\!\!\!\!\!\!\!\!\! \exp\left[\sum_{N=0}^\infty \frac{1}{N!}\left\langle\left(-i\, \kperp\cdot \left(\sperp-\sperpo\right)-i\, \kparal\cdot\left(\sparal-\sparalo+\frac{\dsparal-\dsparalo}{H}\right)\right)^N \right\rangle\right].
\eea

We are interested in resumming only the linear part of the displacement, so that we can define
\bea
&&\!\!\!\!\!\!\!\!\!\!\!\! K_{r,0}(\kvec, \qvec ;t)= \label{k0r} \\ \nonumber
&&\!\!\!\!\!\!\!\!\!\!\!\! \exp\left[- \frac{1}{2}\left\langle\left( \kperp\cdot \left(\sperp_1-\sperpo_1\right)+ \kparal\cdot\left(\sparal_1-\sparalo_1+\frac{\dsparal_1-\dsparalo_1}{H}\right)\right)^2 \right\rangle\right].
\eea
where the subscript $_1$ indicates the linear solution.\footnote{This construction can be generalized to resumming non-linear displacement. See~\cite{Senatore:2014via} for details.}  To simplify this expression further, we use the fact that $\dot {\vec s}_1=H f \vec s_1$, with $f=\d\log D/\d\log a$ and $D$ the growth factor.\footnote{This is true only to the extent that the time dependence can be approximated by replacing factors of $a$ in the EdS solution with factors of the growth factor $D$.  This has been checked to be accurate to percent level.  Any mistake made due to this is parametrized by $\tilde{\epsilon}_{s<}$, which is smaller than unity and will be recovered order by order in perturbation theory~\cite{Senatore:2014via}. }  We can write
\bea \label{k0}
&& K_{r,0}(\kvec , \qvec ;t) =\exp\left[- \frac{1}{2}\left\langle\left(\vec{\tilde k} \cdot \left(\vec s(\vec q,t)_1 - \vec s(\vec 0,t)_1  \right)   \right)^2 \right\rangle\right]\ ,
\eea
with 
\be
\vec{\tilde{k}} \equiv \kvec + f k ( \kdotz ) \zhat  \ . 
\ee
Thus, $K_{r,0}$ is the same function of $\vec{\tilde{k}}$ that $K_0$ (from \cite{Senatore:2014via}) is of $\kvec$: 
\be
K_{r,0}\left(\kvec , \qvec ;t\right)=K_{0}\left(\tkvec (\kvec ),\qvec;t\right)\ .
\ee
We now proceed by defining 
\be
F_{||_{N-j}}   \left(\vec{\tilde k} (\kvec),\qvec  ;   t       \right)   =    K_{0}\left(\vec{\tilde k} (\kvec) , \qvec  ; t   \right) \cdot\left.\left.K_{0}^{-1}\left(\vec{\tilde k} (\kvec),\qvec ; t\right)\right|\right|_{N-j} \ ,
\ee
where the notation $g ||_{i}$ means to expand the quantity $g$ {\it up to} order $i$ in all of $\epssm$, $\epssM$, and $\epsdm$.  Doing this to \eqn{cumulant} would give the standard Eulerian perturbative expansion.  

For the rest of our manipulations, it will be convenient to consider the power spectrum as a function of $k$ and $\khat \cdot \zhat \equiv \mu_k$ instead of $k_\parallel$ and $k_\perp$.  Often we will use the notation $\mu_V \equiv \hat{V} \cdot \zhat$ to define the cosine of the angle between a given vector $\vec{V}$ and the $z$-axis.  Using the notation $g|_i$ to mean that we expand the quantity $g$ \emph{up to} order $i$ in $\epsilon_{\delta <}  $ and $\epsilon_{s >}$, but keep $\epsilon_{s <}$ to all orders (i.e. resum $\epsilon_{s<}$), we can finally write the resummed power spectrum as
\begin{align} \label{psrs}
P^r( k, \khat \cdot \zhat ; t ) \Big|_N & = \int d^3 q'  \,  e^{- i \kvec \cdot \qvecp} \sum_{j = 0}^{N} \left[ F_{||_{N-j}} \left( \vec{\tilde{k}}  , \qvecp ; t \right)  \cdot K_r( \kvec , \qvecp ; t )_j   \right]  \ .
\end{align} 
The function $K_r( \kvec , \qvecp ; t )_j $ is related to the $j$th order piece of the Eulerian power spectrum by
\be
P^r ( k , \kdotz ; t )_j  = \int d^3 q \, e^{- i \kvec \cdot \qvec} \, K_r ( \kvec , \qvec ; t)_j \ .
\ee
Next we evaluate \eqn{k0} following Appendix A of \cite{Senatore:2014via} to get
\begin{align} \nonumber
K_0 ( \vec{ \tilde{ k}} , \qvec ; t) &  = \exp \left\{ - \half  \left( X_1 ( q ; t  ) \delta_{ij} + Y_1(q ; t ) \hat q_i \hat q_j \right) \hat{\tilde{k}}^i \hat{\tilde{k}}^j \right\}  \\ \nonumber
& = \exp \Bigg\{ - \frac{k^2}{2} \Big[ X_1( q ; t) \left( 1 + 2 f \mu_k^2 + f^2 \mu_k^2 \right) \\
& \hspace{1in} + Y_1 ( q ; t) \left(  (\kdotq)^2 + 2 f \mu_k \mu_q \kdotq + f^2 \mu_k^2 \mu_q^2 \right) \Big] \Bigg\} \ , 
\end{align}
where 
\bea \nonumber
&& X_1(q;t)=\frac{1}{2\pi^2}\int_0^{+\infty} dk\;  \exp\left[- \frac{k^2}{\Lambda_{\rm IR}^2}\right]\;P_{11}(k;t) \left[\frac{2}{3}-2\,\frac{j_1(k q)}{k q}\right]\  \\
&& Y_1(q;t)=\frac{1}{2\pi^2}\int_0^{+\infty} dk\;\exp\left[- \frac{k^2}{\Lambda_{\rm IR}^2}\right]\; P_{11}(k;t) \left[- 2\, j_0(k q)+6\,\frac{j_1(k q)}{k q}\right]\ 
\eea
and $j_0$ and $j_1$ are the zeroth and first spherical Bessel functions respectively.  Here, $\Lambda_{\rm IR}$ is the IR scale up to which we resum the linear IR modes.  This cutoff and the choice in \eqn{k0r} to keep the linear modes non-perturbative both serve to define a new expansion parameter $\tilde \epsilon_{s<} $, such that $\tilde \epsilon_{s<} \ll 1 \lesssim \epsilon_{s<}$, in terms of which the perturbative expansion of the IR displacements is now done.  Taking $\Lambda_{\rm IR}$ too high would mean that we include some uncontrolled UV modes, but this mistake would be at the next order in $\epsilon_{\delta <}$ and would be recovered order by order in the loop expansion.  In practice, we use $\Lambda_{\rm IR} = 0.066 \unitsk$.  Now, we proceed in a way similar to the real space calculation 
\begin{align} \nonumber
 P( k, \khat \cdot \zhat ; t) \Big|_N  & = \left( \int_{\kvec'} (2 \pi)^3 \delta( \kvec - \kvec') \right) \int d^3 q'  \,  e^{- i \kvec \cdot \qvecp} \sum_{j = 0}^{N} \left[ F_{||_{N-j}} ( \tkvec, \qvecp; t )  \cdot K_r( \kvec , \qvecp; t)_j   \right] \\ \nonumber
& = \left( \int_{\kvec'} (2 \pi)^3 \delta( \kvec - \kvec') \right) \int d^3 q'  \,   \sum_{j = 0}^{N} \left[ F_{||_{N-j}} ( \tkvec, \qvecp  ; t )  \cdot e^{- i \kvec' \cdot \qvecp}K_r( \kvec' , \qvecp; t)_j   \right] \\  \nonumber
& = \left( \int_{\kvec'}  \int d^3 \qvec \, e^{i \qvec \cdot ( \kvec' - \kvec)} \right) \int d^3 q'  \,  \sum_{j = 0}^{N} \left[ F_{||_{N-j}} ( \tkvec , \qvecp ; t)  \cdot e^{- i \kvec' \cdot \qvecp}K_r( \kvec' , \qvecp; t)_j   \right] \\  \nonumber
& =  \sum_{j = 0}^{N}  \int_{\kvec'}  \int d^3 q \, e^{i \qvec \cdot ( \kvec' - \kvec)}  F_{||_{N-j}} ( \tkvec , \qvec ; t) \int d^3 q'  \,   e^{- i \kvec' \cdot \qvecp}K_r( \kvec' , \qvecp ; t)_j    \\   
& =  \sum_{j = 0}^{N}  \int_{\kvec'}  \int d^3 q \, e^{i \qvec \cdot ( \kvec' - \kvec)}  F_{||_{N-j}} ( \tkvec , \qvec ; t ) \, P^r(k' , \hat k ' \cdot \zhat ; t)_j  \ , \label{manip}
\end{align} 
where we use the notation $\int_{\kvec} \equiv \int \momspmeas{k}$.  The trick in this passage is replacing $F_{||_{N-j}} ( \tkvec , \qvecp ; t) \rightarrow F_{||_{N-j}} ( \tkvec , \qvec ; t)$ between the third and fourth lines, which amounts to making a mistake proportional to the gradients of the displacements, which is higher order in $\epsilon_{\delta <}$ than we take into account at order $N$.

Now we will expand the above expression in multipoles using the Legendre decomposition in \eqn{legendre} to get (dropping the $t$ dependence to remove clutter)
\begin{align} \label{presum1}
P_\ell^r ( k ) \Big|_N & = \sum_{j = 0}^N \sum_{\ell'} \int \frac{d k' \, k'^2}{2 \pi^2} M_{||_{N-j}} ( k , k')_{\ell \ell'} P^r_{\ell'} ( k' )_j \\
M_{||_{N-j}} ( k , k')_{\ell \ell'} & = \int d q \,\,j_{\ell ' } ( k' q ) \,  Q_{||_{N-j}} ( k , q )_{\ell \ell'} \label{sbt} \\
Q_{||_{N-j}} ( k , q )_{\ell \ell'} & =  i^{\ell '} \, q^2 \frac{2 \ell + 1}{2} \int_{-1}^1 d \mu_k  \int d^2 \hat q \, e^{- i \qvec \cdot \kvec }     F_{||_{N-j}} ( \vec{\tilde{k}} , \qvec  )     \,        \cp_\ell ( \mu_k)  \, \cp_{\ell '} ( \mu_q )   \label{qmxbad}
\end{align}
where $j_\ell ( x)$ is the spherical Bessel function of type $\ell$.  For the one-loop calculation that interests us here, $\ell$ and $\ell'$ take on the values $0,2,4,6$, and $8$.  The computational strategy appears to be to first compute the $Q$ functions, and then use a spherical Bessel transform to evaluate the $M$ functions.  Compared to the real space resummation, there are three extra angular integrals to perform here for each $M_{\ell \ell'}$, and we must compute five values of $\ell'$ for each $\ell$.  This makes a brute force numerical strategy challenging, so we will simplify the calculation below.

\subsection{Evaluation of Q}

The main trick that we use is to expand $F_{||_{N-j}}$ in powers of $k^2 Y_1(q)$.  This is justified because $k^2 Y_1 ( q ) $ times the angular factors is small over most of the angular integration.\footnote{We are helped by the fact that, even for order 1 arguments, the exponential quickly becomes numerically close to its expansion.  For example, the difference between $e^x$ and its expansion to order $x^3$ at $x=1$ is 2\%.  This is due to the infinite radius of convergence of $e^x$.  For us, at the scales of interest $k \approx 0.2 \unitsk$ and $q = q_{BAO} \approx 105\, h^{-1} {\rm Mpc}$, and taking an average value of $\mu^2 = 1/3$ for the angular factors, $k^2 Y_1(q) \mu^2 / 2 = 1.37$.}  Once we do this, all of the integrals in \eqn{qmxbad} can be done analytically and the numerical evaluation is much quicker.   To proceed, we write 
\begin{align} \nonumber\label{eq:temp}
K_0 ( \vec{\tilde{k}} , \qvec  ) = & \,  e^{-\frac{k^2}{2} \left( X_1(q) + \alpha Y_1 ( q) \right) }   e^{- \frac{k^2}{2} X_1(q) \mu_k^2 f \left( 2 + f \right) } \times  \\
& \hspace{.3in} \exp \left\{ - \frac{k^2}{2} Y_1 ( q ) \left( ( \kdotq)^2 - \alpha + 2 f \mu_k \mu_q \kdotq + f^2 \mu_k^2 \mu_q^2 \right) \right\} \ . 
\end{align}
The first two factors above will be treated analytically, so it will be convenient to separately define the second line of (\ref{eq:temp}) as 
\be  \bar K_0  ( \vec{\tilde{k}} , \qvec  )     = \exp \left\{ - \frac{k^2}{2} Y_1 ( q ) \left( ( \kdotq)^2 - \alpha + 2 f \mu_k \mu_q \kdotq + f^2 \mu_k^2 \mu_q^2 \right) \right\}\ ,
\ee
and expand $  \bar K_0  ( \vec{\tilde{k}} , \qvec  )  \cdot K_0^{-1} ( \vec{ \tilde{ k}} , \qvec ) \Big| \Big|_{N-j}  $ to the desired order $n_Y$ in $k^2 Y_1$.  It turns out that first order is sufficient for our purposes, but the method is general, and expanding to higher orders does not cost much extra computational time.  In Appendix~\ref{ircompare}, we run some convergence checks on this approximation.  We also introduced a number $\alpha$, which can be chosen to make the value of the angular terms multiplying $Y_1$ smaller so that the expansion converges more quickly.  In practice, we choose $\alpha = 1/3$, which is the average of $(\kdotq)^2$ over the angular integrals in Q.  After expanding in $k^2 Y_1$, it is clear that the general integral that we have to evaluate is 
\begin{align} \label{lint}
L^{abc}_{\ell \ell' } ( k , q ) \equiv \frac{i^{\ell ' }}{ 4 \pi } \int_{-1}^1 d \mu_k \int d^2 \hat q \, \cp_\ell ( \mu_k ) \cp_{\ell '} ( \mu_q ) e^{ - i \kvec \cdot \qvec} e^{ - \frac{k^2}{2} X_1(q) \mu_k^2 f \left( 2 + f \right) }\,  \mu_k^a \, \mu_q^b \, (\kdotq)^c
\end{align}
for various pawers of $a,b,c$, and where the prefactor has been chosen with foresight.  Then, to write \eqn{qmxbad} in terms of \eqn{lint}, first extract the coefficients of the angles by expanding 
\be  \label{qmxgood}
 \bar K_0  ( \vec{\tilde{k}} , \qvec  )  \cdot K_0^{-1} ( \vec{ \tilde{ k}} , \qvec ) \Big| \Big|_{N-j} = \left[ \kappa_{||_{N-j}}^{(n_Y)} (k , q) \right]_{abc}\, \mu_k^a \, \mu_q^b \, ( \kdotq )^c
 \ee
where $ [ \kappa_{||_{N-j}}^{(n_Y)} (k , q) ]_{abc}$ is expanded up to order $n_Y$ in $k^2 Y_1$,\footnote{ While the factor $ [ \kappa_{||_{N-j}}^{(n_Y)} (k , q) ]_{abc}$ contains only up to order $n_Y$ in $Y$, we will eventually multiply it by the order $j$ Eulerian power spectrum, which contains terms at order $j$ in $Y$.  Thus, in particular, if we are resumming the $N$ loop power spectrum, our resummation expression contains terms up to order $n_Y + N$ in $Y$.  Our resummation scheme ensures that, if one chooses $n_Y < N$, the terms with $Y$ at the order between $n_Y$ and $N$ are simply the terms from Eulerian perturbation theory.  Note also that $n_Y = N$ is not a trivial resummation (as one might expect since we are getting correct the terms up to $Y^N$, which indeed are just the Eulerian terms).  This is because our expansion also captures correctly terms up to $Y^{n_Y} \epsilon_{\delta <}^N$, which indeed are \emph{not} captured by the order $N$ Eulerian expansion.  Even  more generally, our expansion scheme converges because we are approximating the $M_{||_{N-j}}$ matrices with a fixed order truncation of their convergent analytic expansion (as they depend analytically on $Y$). Of course, this is a good procedure once the cancellation of the relevant terms among the different $M_{||_{N-j}}$ matrixes is ensured, as in this case.   } and a summation over $a,b,c$ is implied.  In terms of these expressions, the final answer is, up to order $n_Y$ in $k^2 Y_1$,  
\begin{align}\label{eq:temp3}
Q_{||_{N-j}} ( k , q )_{\ell \ell'} = 2 \pi q^2 ( 2 \ell  + 1) e^{-\frac{k^2}{2} \left( X_1(q) + \alpha Y_1 ( q) \right) }   \left[ \kappa_{||_{N-j}}^{(n_Y)} (k , q) \right]_{abc}L^{abc}_{\ell \ell'} (k , q) \ . 
\end{align}

Now we turn to the evaluation of $L^{abc}_{\ell \ell '}$.  First, write 
\be\label{eq:temp1}
\mu_q^b \, \cp_{\ell '} ( \mu_q ) = \sum_{j = -b}^b n^b_j ( \ell ') \cp_{\ell' + j } ( \mu_q)
\ee
where 
\be
n^b_j ( \ell ') = \begin{cases} \frac{ 2( \ell ' + j ) + 1 }{2} \int_{-1}^1d \mu  \, \mu^b \cp_{\ell' + j } ( \mu ) \cp_{\ell '} ( \mu ) & \text{ for} \hspace{.2in} \ell' +j \geq 0 \\ 0 & \text{ for} \hspace{.2in} \ell' + j < 0 \end{cases} \ . 
\ee
This kind of expansion should be familiar from the angular momentum algebra.  Next, using the Legendre decomposition of a plane wave from \eqn{planewave}, notice that 
\be\label{eq:temp2}
( \kdotq )^c e^{- i k \, q \, \kdotq} = i^c \frac{\partial^c}{\partial x^c } e^{- i x\, \kdotq} \Bigg|_{x = k q} = i^c \sum_{\tilde \ell} (-i)^{\tilde \ell} ( 2 \tilde \ell + 1) j_{\tilde \ell}^{(c)} ( k q ) \cp_{\tilde \ell} ( \kdotq) \ 
\ee
where $j_{ \ell}^{(c)} ( k q ) = \frac{\partial^c}{\partial x^c} j_\ell ( x ) \Big|_{x = k q}$.  After plugging (\ref{eq:temp1}) and (\ref{eq:temp2}) into \eqn{lint}, we can do the integral over $d^2 \hat q$ using \eqn{legpoly1} to get 
\be \label{leq}
L^{abc}_{\ell \ell' } ( k , q ) = i^c \sum_{j = - b}^b (-i)^j \,   n_j^b ( \ell ' ) \,  j_{\ell' + j}^{(c)} ( k q)\, I^a_{\ell \ell' + j} ( k , q)
\ee
where
\be \label{ieq}
I^a_{\ell \ell' } ( k , q) = \int_{-1}^1 d \mu_k \,  \cp_\ell ( \mu_k )\,  \cp_{\ell' }( \mu_k ) \, \mu^a_k \, e^{ - \frac{k^2}{2} X_1(q) \mu_k^2 f \left( 2 + f \right) } \ . 
\ee 
The last integral can be done for general $a$ in terms of Erfi functions.  However, since we are interested in doing this computation quickly numerically, and since the argument of the exponential is small on the scales of interest, we can Taylor expand the exponential to a small order, and it will converge.  Thus, we actually use  
\be
I^a_{\ell \ell' } ( k , q) \approx e^{ - \frac{k^2}{2} X_1(q) \beta f \left( 2 + f \right) }    \int_{-1}^1 d \mu_k \cp_\ell ( \mu_k ) \cp_{\ell' }( \mu_k ) \mu^a_k \, e^{ - \lambda \frac{k^2}{2} X_1(q)\left(  \mu_k^2 - \beta \right) f \left( 2 + f \right) } 
\ee
where we first expand $\lambda$ to the desired order $n_X$, set $\lambda = 1$, do the integral over the polynomials in $\mu_k$, and choose $\beta = 1/3$ to make the expansion converge more quickly.  In practice, we use $n_X = 3$ and $n_Y = 1$.  In this way, we have Mathematica precompute $L^{abc}_{\ell \ell'} ( k , q ) $ for each $a,b,c,\ell,\ell'$ analytically in $k$ and $q$, and so the resulting expressions can be used for any cosmology (i.e. for any $X_1$ and $Y_1$).  Then, we evaluate $Q_{||_{N-j}} ( k , q )_{\ell \ell'}$ in \eqn{eq:temp3} at discrete $k$ and $q$ points to do the spherical Bessel transform.  Once we have the expressions analytically in terms of $k$ and $q$, the evaluation at discrete points is very quick.  Roughly speaking, it is about $600$ times faster to compute all of the $(k,q)$ points (about $33,000$ points after the numerical improvement we discuss next) in $Q_{||_{1}} ( k , q )_{\ell \ell'}$ for a given $(\ell, \ell')$ by our \eqn{eq:temp3} than by the full expression \eqn{qmxbad}.  Of course, we first have to compute the various terms in \eqn{leq}, the longest of which is \eqn{ieq}, but this is negligible because it only needs to be done once for each pair $(\ell , \ell')$.

\subsection{A Numerical Improvement}
In order to stabilize the $q \rightarrow \infty$ behavior of $Q_{||_{N-j}} ( k , q )_{\ell \ell'}$ in the spherical Bessel transform in \eqn{sbt}, consider the limit $X_1(q) \rightarrow X_1(  \infty) \equiv X_1^\infty$ and $Y_1 ( q ) \rightarrow Y_1 (  \infty ) \equiv Y_1^\infty = 0$.  Since $Y_1^\infty = 0$, the components of $ \left[ \kappa_{||_{N-j}}^{(n_Y)} (k , q) \right]_{abc}$ with $b\neq0$ and $c\neq 0$ are all zero.  This leads us to consider 
\be
Q^\infty_{||_{N-j}} ( k , q )_{\ell \ell'} = 2 \pi q^2 ( 2 \ell  + 1) e^{-\frac{k^2}{2}  X_1^\infty   }   \left[ \kappa_{||_{N-j}}^{(n_Y)} (k , \infty) \right]_{a00}    j_{\ell ' } ( k q ) I^a_{\ell \ell'} ( k , \infty) \ .
\ee
Then, what we actually compute is (adding and subtracting \eqn{presum1} but with $Q^\infty$ instead of $Q$)
\begin{align} \nonumber
P_\ell^r ( k ) \Big|_N & = \sum_{j = 0}^N \sum_{\ell'} \Bigg( \int \frac{d k' \, k'^2}{2 \pi^2} \Delta M_{||_{N-j}} ( k , k')_{\ell \ell'} P^r_{\ell'} ( k' )_j  \\
&\hspace{.5in} +  \frac{2 \ell + 1}{2} e^{-\frac{k^2}{2} X_1^\infty }  \left[ \kappa_{||_{N-j}}^{(n_Y)} (k , \infty) \right]_{a00}   I^a_{\ell \ell'} ( k , \infty) P^r_{\ell'} ( k ) _j \Bigg)
\end{align}
where
\begin{align} 
\Delta M_{||_{N-j}} ( k , k')_{\ell \ell'} & = \int d q \,\,j_{\ell ' } ( k' q ) \,  \Delta Q_{||_{N-j}} ( k , q )_{\ell \ell'} \label{sbt2} \\
\Delta Q_{||_{N-j}} ( k , q )_{\ell \ell'} & =  Q_{||_{N-j}} ( k , q )_{\ell \ell'}  - Q^\infty_{||_{N-j}} ( k , q )_{\ell \ell'}  \ . 
\end{align}
Analytically, this is equivalent to \eqn{presum1}, but the integral in \eqn{sbt2} is better behaved than the one in \eqn{sbt}.  This means that we need to evaluate $\Delta Q_{||_{N-j}} ( k , q )_{\ell \ell'}$ at fewer $q$ points to achieve the desired precision in the spherical Bessel transform.

\section{Comparison to Numerical Simulations \label{sims}}

In this section, we compare our results to the large BigMultiDark $N$-body numerical simulation at $z=0.56$ with a volume of $(2.5 {\rm Gpc}/h)^3$ and $3840^3$ particles (see~\cite{Klypin:2014kpa} for details), which uses the Planck $\Lambda$CDM cosmology: $\left\{ \Omega_m , \Omega_b, \Omega_k, H_0,\sigma_8, n_s \right\}$ $= \left\{ 0.307 , 0.048 , 0, 67.77,0.96 \right\}$.  Results at $z=0$ are presented in Appendix \ref{z0}.  First, we compare to the dark-matter power spectrum, to determine the value of $c_s^2$.  To do the resummation in real space, we used a method similar to the one described above.  In Appendix~\ref{ircompare}, we run some tests on the new method of IR-resummation for the real space, dark-matter power spectrum.  Also, we will ignore some of the subtleties of overfitting described in \cite{Foreman:2015lca} (which describes in detail and performs a rigorous fitting procedure for the dark matter power spectrum) for simplicity in this paper; in reality, a more involved analysis is necessary to do an accurate fit.  In this vein, we simply identify some values of the parameters such that the EFT curve is compatible with the data; in this paper, we are most interested in developing the correct formalism and checking its consistency in redshift space rather than performing an accurate fit.  That said, as a very rough estimate of the error, we noticed that if one shifts  $c_s^2$ by about $5\%$, $\bar c_1^2$ by about $15\%$, or $\bar c_2^2$ by about $50\%$, the theory fails at a much lower value of $k$ than if one uses the values that we use in the plots.

 \begin{figure}[htb!] 
\begin{center}
\includegraphics[width=10cm]{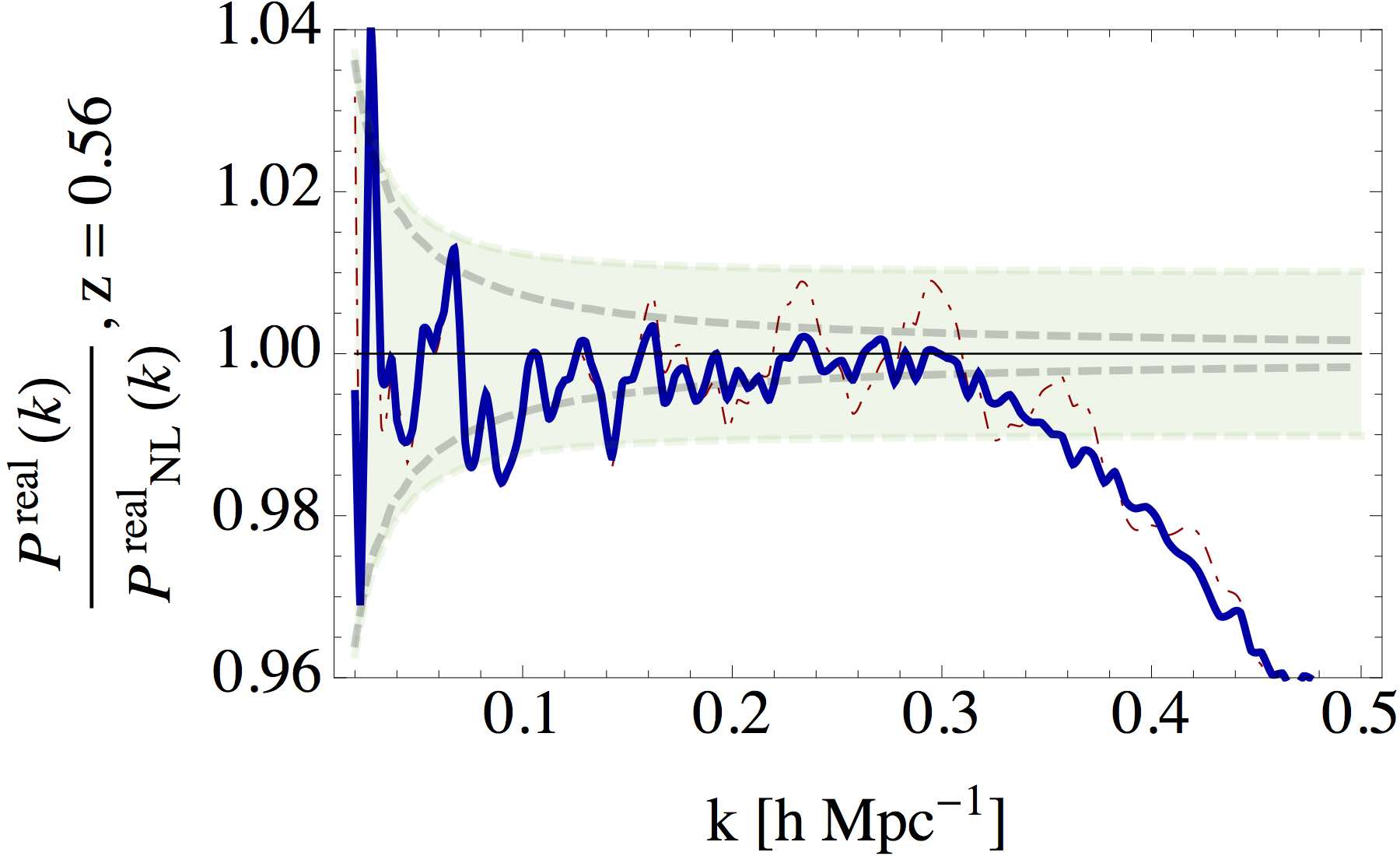} 
\caption{  Here, we compare the EFT prediction with with $c_s^2  = 0.31  \, (\knl/ \unitsk)^2$ to the BigMultiDark dark-matter power spectrum~\cite{Klypin:2014kpa} at $z=0.56$ .  The dark blue curve is the one-loop IR-resummed EFT prediction, the red dot-dashed curve is the non-resummed EFT prediction, the gray dot-dashed curve is one-loop SPT, and the green region is the error on the data, where we have included a $1\%$ systematic error.  We have also plotted the cosmic variance separately as a dashed grey curve to show that this largely explains the fluctuations in the data. }  \label{dmz56}
\end{center}
\end{figure}

 In Figure~\ref{dmz56} we compare with the dark-matter power spectrum in real space at $z=0.56$, and in Figure~\ref{pls} and Figure~\ref{pls2}, we compare with the redshift space multipoles.  In Figure~\ref{pls} and Figure~\ref{pls2}, we use the values
\begin{align}
c_s^2 ( 0.56 ) & = 0.31  \, (\knl / \unitsk)^2 \\
\bar c_1^2 ( 0.56 )& = -3.47 \,  (\knlv / \unitsk)^2 \\
\bar c_2^2 (0.56) & = 0.81 \, ( \knlv / \unitsk) ^2 \ .
\end{align}  
With these parameters, the theory fits the data until $k \approx 0.13 \unitsk$, where the one-loop SPT calculation fails around $k \approx 0.05 \unitsk$.  Notice that $\bar c_1^2$, the coupling associated with the redshift space counterterms, is much larger than the dark matter coupling $c_s^2$.  This suggests that the nonlinear scale associated with the redshift space counter terms is actually significantly smaller than the one for dark matter: $\knlv \approx  \knl / 3$.  This is discussed in more detail in Section~\ref{smallkv}.  We do not show the $\ell = 8$ multipole because it is very small with large errors.

Figure~\ref{pls2} can also be used to have a sense of the systematic error in the numerical data that we use, as it is expectable than nature provides a relatively smooth curve. We see that a percent level comparison of theory and data is made difficult by this situations.\footnote{To have a better sense of the numerical systematic error, we have asked several people in that community for additional numerical data. Unfortunately these are the only (and therefore the best) data we were able to find. If the interested reader is able to provide us with more accurate data, we will be happy to include a comparison with them in this paper.}

\begin{figure}[htb!] 
\begin{center}
\includegraphics[width=8cm]{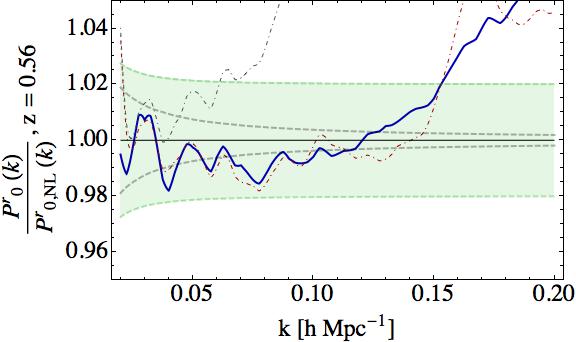} \includegraphics[width=8cm]{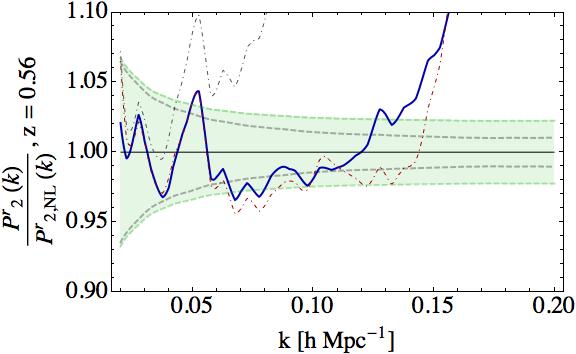} \\
\includegraphics[width=8cm]{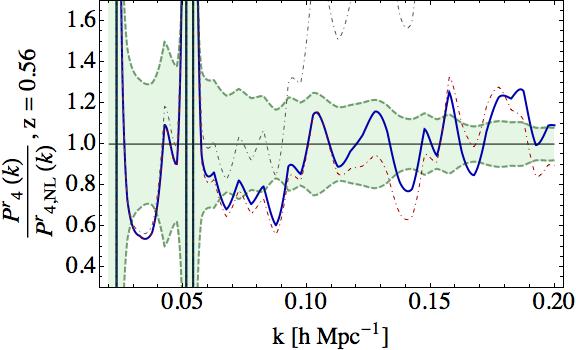} \includegraphics[width=8cm]{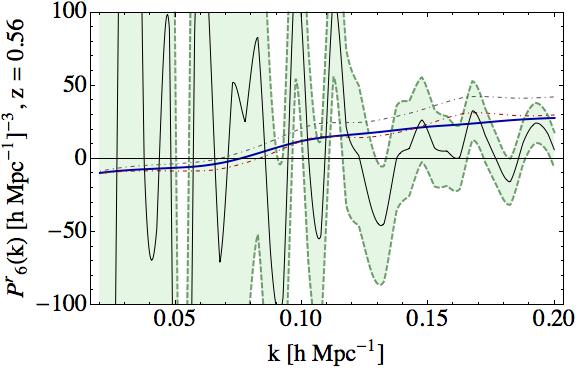}
\caption{  Here, we present the results in redshift space at $z=0.56$ for $\ell = 0,2,4,6$, with $\bar c_1^2=-3.47 \,  (\knlv / \unitsk)^2$ and $\bar c_2^2 = 0.81 \, ( \knlv  / \unitsk) ^2$.  The dark blue curve is the one-loop IR-resummed EFT prediction, the dot-dashed red curve is the non-resummed EFT prediction, the gray dot-dashed curve is one-loop SPT, and the green region is the error in the data, where we have included a $2\%$ systematic error.  The plots fit until about $k \approx 0.13 \unitsk$.  In the last plot, the non-linear data is very noisy, so we do not present the ratio of power spectra.  However, it is still clear that the prediction is within the errors.  We have also plotted the cosmic variance separately as a dashed grey curve to show that this largely explains the fluctuations in the data.  Note that for $\ell = 4$ and $\ell = 6$ the error is completely dominated by the cosmic variance.}  \label{pls}
\end{center}
\end{figure}

\begin{figure}[htb!] 
\begin{center}
\includegraphics[width=8cm]{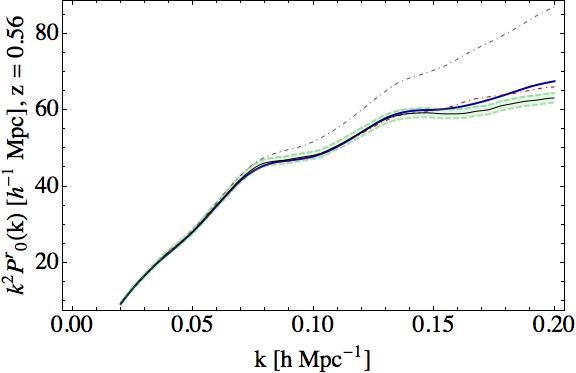} \includegraphics[width=8cm]{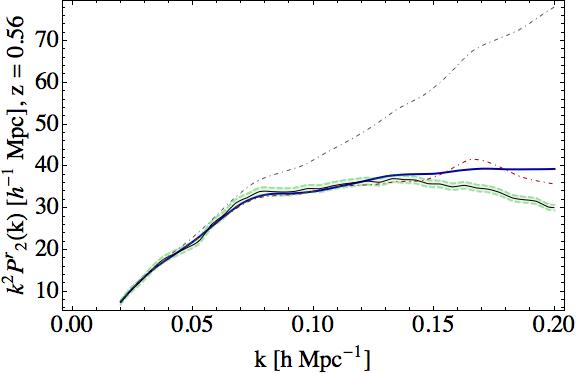} \\
\includegraphics[width=8cm]{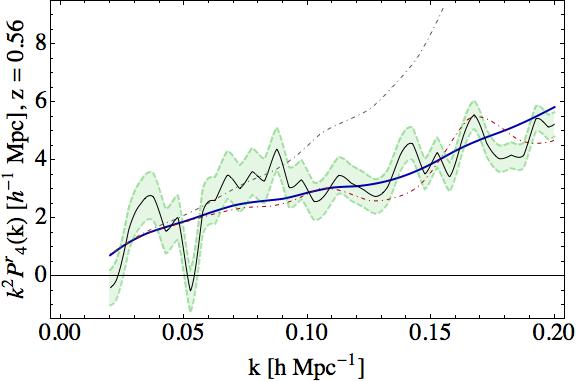} \includegraphics[width=8cm]{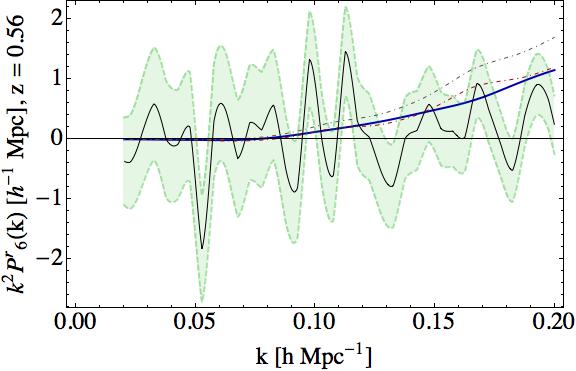}
\caption{  Here, we present the results in redshift space at $z=0.56$ for $\ell = 0,2,4,6$, with $\bar c_1^2=-3.47 \,  (\knlv / \unitsk)^2$ and $\bar c_2^2 = 0.81 \, ( \knlv  / \unitsk) ^2$, the same as in Figure~\ref{pls}, but we plot $k^2 P^r_\ell (k)$.    The dark blue curve is the one-loop IR-resummed EFT prediction, the dot-dashed red curve is the non-resummed EFT prediction, the gray dot-dashed curve is one-loop SPT, the thin black line is the data, and the green region is the error in the data, where we have included a $2\%$ systematic error, which, for the higher multipoles, is probably an underestimate (see text for additional comments).  The plots fit until about $k \approx 0.13 \unitsk$. }  \label{pls2}
\end{center}
\end{figure}

In Figure \ref{resum} we also include the result of the IR-resummation up to $k = 0.5 \unitsk$ to show the success of the method described in this paper.  The resummation should capture \emph{all} of the BAO oscillations, even those which are outside of the UV reach of the theory, without changing the reach of the theory.  This idea has been discussed in many previous works~\cite{Scoccimarro:1995if,Carrasco:2013sva,Creminelli:2013mca}.  Because the IR-resummation only effects IR modes, the equivalence principle demands that the UV properties of the theory cannot be affected \cite{Carrasco:2013sva}.  Thus, in the UV, the resummed and non-resummed curves should be the same apart from the approximation that we made in \eqn{manip} that is higher order in $\epsilon_{\delta <}$.  Indeed, this is the case for all of the multipoles.  In Figure \ref{resum}, we have chosen the values of the speed of sound parameters to emphasize the result, not to correspond to the physical theory.

\begin{figure}[htb!] 
\begin{center}
\includegraphics[width=8cm]{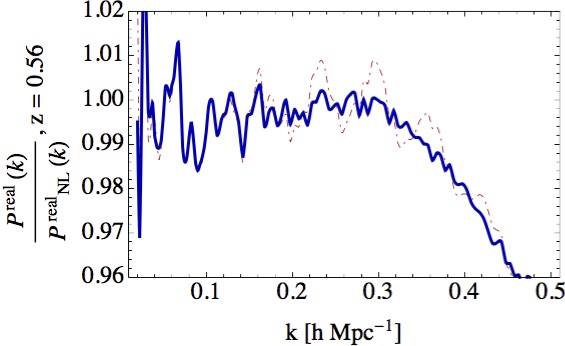} \includegraphics[width=8cm]{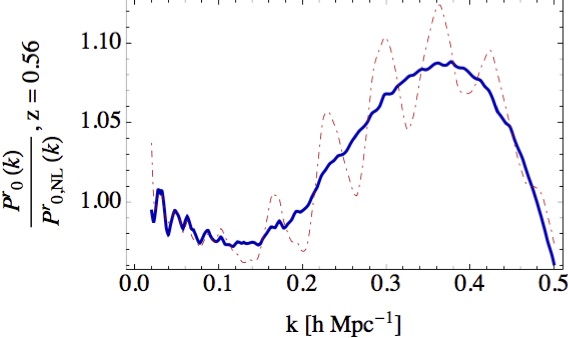} \\
\includegraphics[width=7.9cm]{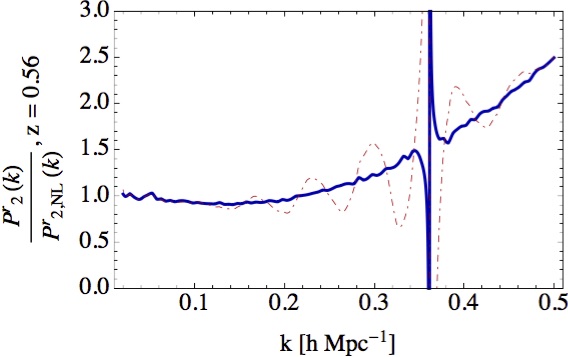} \includegraphics[width=8cm]{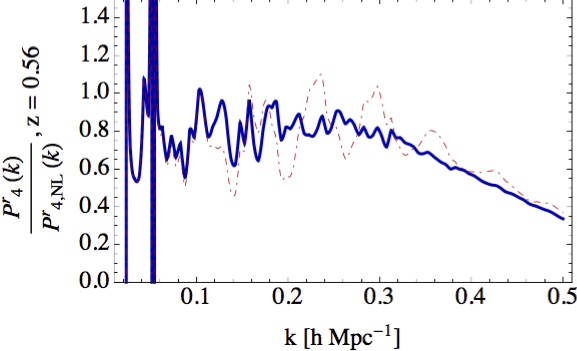}
\caption{  Here, we present the resummation results in redshift space at $z=0.56$ for the real-space power spectrum and $\ell = 0,2,4$.  The dark blue curve is the one-loop IR-resummed EFT prediction and the dot-dashed red curve is the non-resummed EFT prediction.  The values of the speed of sound parameters have been chosen so that the curves only change by about $10\%$  up to $k = 0.5 \unitsk$ in order to emphasize the result of the resummation.  As expected, the resummation does not change the UV part of the theory; it just captures the BAO oscillations. }  \label{resum}
\end{center}
\end{figure}

\section{Small $\knlv$ Limit  \label{smallkv}}

As mentioned earlier, it is reasonable to expect that $\knlv < \knl$ because velocities become non-linear before over-densities.  This is also suggested by the data, since $\bar{c}_1^2$ is much larger than $c_s^2$.  If this is the case, we are justified in adding extra counter terms, such as $( k / \knlv)^4P_{11}$ or potentially even the stochastic terms proportional to $(k / \knlv)^4$ (as we will see below), which can be larger than the dark matter two-loop contributions.  To estimate when this is valid, we can approximate the behavior of $P_{11}$ by~\cite{Carrasco:2013mua}
\be
P_{11}(k) =  (2\pi)^3 \begin{cases}  \frac{1}{ \knl^3 } \left( \frac{ k}{  \knl } \right)^{n} & \text{for} \, \,  \, k < k_{\rm tr}  \\    \frac{1}{ \tilde{k}_{\rm NL}^3 } \left( \frac{ k}{  \tilde{k}_{\rm NL} } \right)^{\tilde{n}}  & \text{for} \, \, \, k > k_{\rm tr}   \end{cases}
\ee
with $\tilde{k}_{\rm NL} = 3.96 \unitsk$, $\knl = 1.25 \unitsk$, $\tilde n = -2.11$, $ n  = -1.54$, and $k_{\rm tr} =0.21 \unitsk$.  Then, consider the leading two-loop piece 
\be
\left( \frac{k }{\knlv} \right)^2 P_{1-\rm loop} \sim   \left( \frac{k }{\knlv} \right)^2 \left( \frac{k }{\knl} \right)^{3 + n} P_{11}
\ee
where $n \approx -1.5$ is the slope of the linear power spectrum.  Then the new counter terms  $( k / \knlv)^4P_{11}$ dominate if 
\be
 \left( \frac{k }{\knlv} \right)^4 \gtrsim \left( \frac{k }{\knlv} \right)^2 \left( \frac{k }{\knl} \right)^{3 + n}  \ . 
 \ee
We want these terms to dominate for $k \approx 0.2 \unitsk$, which means we need
\be
\left(0.2 \unitsk\right)^{0.5} (\knl)^{1.5} > (\knlv)^2
\ee
which for $\knl \approx 1.25 \unitsk$, means we need $\knlv \approx 0.79 \unitsk$, which seems reasonable given the parameters that we found earlier.  The stochastic terms will be important when 
\begin{align} \nonumber
\frac{1}{k_{\rm NL}^{r\,3}} \left( \frac{k }{\knlv} \right)^4  & \gtrsim \left( \frac{k }{\knlv} \right)^2 \left( \frac{k }{\knl} \right)^{3 + n}  P_{11}   \\
 & \sim \left( \frac{k }{\knlv} \right)^2 \left( \frac{k }{\knl} \right)^{3 + n}  \frac{(2 \pi )^3}{ k_{\rm NL}^3 } \left( \frac{ k}{  k_{\rm NL} } \right)^{ n} 
\end{align}
which means $\knlv \approx 0.2 \unitsk$.  This is a much smaller scale than the one above, so we anticipate that the stochastic terms will not contribute significantly.  However, for the sake of this analysis, we assume that $\knlv$ is small enough to warrant the inclusion of the stochastic term.  In this case, in addition to the stochastic pieces, the new counter terms that we should look for are the higher derivative terms linear in the fields.  These are
\begin{align} 
[v^2_z]_{R , \kvec} & \supset \hat{z}^i \hat{z}^j  \left( \frac{aH}{\knlv} \right)^2   \left\{ c_{\rm st, 1}  \Delta^{ij}_{\rm st, 1}( \kvec) +\left( \frac{k}{\knlv} \right)^2 \left(  c_{1} \delta_{ij} + c_{2} \frac{ k_i k_j}{k^2} \right) \delta(\kvec) \right\}
\end{align}
\begin{align} \nonumber
[v^3_z]_{R,\kvec} & \supset \hat{z}^i \hat{z}^j \hat{z}^l   \left( \frac{a H}{\knlv} \right)^3 \Bigg\{ c_{\rm st, 2}  \Delta^{ijl}_{\rm st, 2}(\kvec)   + c_3 \left( \frac{\knlv}{a H} \right) \left(   \frac{k_i k_j}{k_{\rm NL}^{r \, 2} }  v_l(\kvec)  + 2 \text{  perms.} \right)    + 3 i  c_4     \frac{k_i k_j k_l}{\knlv \, k^2 } \delta(\kvec)     \\
& \hspace{1in} + i c_5   \left( \delta_{ij} \frac{ k_l}{\knlv} + 2 \text{  perms.} \right) \delta(\kvec)       \Bigg\}  \nonumber
\end{align}
\begin{align} \nonumber
[v^2_z \delta ]_{R , \kvec}   & \supset    \hat{z}^i \hat{z}^j  \Bigg\{     \left( \frac{a H}{\knlv} \right)^2  \left( c_{\rm st, 3}  \Delta^{ij}_{\rm st, 3}(\kvec) + c_6 \frac{k_i k_j}{k_{\rm NL}^{r\,2}} \delta(\kvec)  \right)+ i  c_7 \left( \frac{a H}{\knlv} \right) \frac{k^2 (k_i v_j ( \kvec) + k_j v_i ( \kvec))}{k_{\rm NL}^{r\,3}}  \Bigg\}    \ .
\end{align}
We allow the stochastic fields to have any tensor structure that is compatible with the transformation properties of $v^i$. 
At leading order in derivatives, we have 
\be 
\langle \Delta^{ij}_{{\rm st},a} (\kvec ) \Delta^{kl }_{{\rm st},b} ( \kvec ' ) \rangle   = \frac{(2\pi)^3}{(k_{\rm NL}^{r})^3}  \delta_D( \kvec + \kvec' )  \left(   \delta^{ij} \delta^{kl} C^{(1)}_{ab} + \left(\delta^{ik} \delta^{jl}+\delta^{jk} \delta^{il}\right) C^{(2)}_{ab}  \right) \ ,
\ee
and 
\bea
&&\langle \Delta^{ij}_{{\rm st},a} (\kvec ) \Delta^{kl m}_{{\rm st},b} ( \kvec ' ) \rangle  =  \frac{(2\pi)^3}{(k_{\rm NL}^{r})^3}  \delta_D( \vec k + \vec k')  \Bigg(  \left( \delta^{ij} \delta^{kl} \frac{ k^m}{\knlv} + 2 \text{  perms.} \right)C^{(4)}_{ab}    \\  
&& \qquad\qquad   +\left(\left(\delta^{ik} \delta^{jl}+\delta^{jk} \delta^{il}  \right)\frac{ k^m}{\knlv} + 2 \text{  perms.} \right) C^{(5)}_{ab}  +  \left( \frac{k^i}{\knlv} \delta^{jk} \delta^{lm}  + 2 \text{  perms.}    \right)C^{(6)}_{ab}     \Bigg)  \ , \nonumber
\eea
where $C^{(i)}_{ab}$ are order-one numbers. These stochastic terms have no correlations with the matter fields, but they may have correlations with the dark-matter stochastic fields.   

Evaluating these on the linear solutions and combining degenerate terms gives a contribution to $\delta_r ( \kvec ) $ at the next lowest order in $k / \knlv$ of 
\begin{align}
\delta_r ( \kvec ) &  \supset - \half \left( \frac{k}{\knlv} \right)^4 \left( \tilde c_1^2  \mu^2 + \tilde c_2^2  \mu^4 +  \tilde c_3^2  \mu^6 \right) \delta_1 ( \kvec )  - c_{\rm st} \mu^2 \left( \frac{ k }{\knlv} \right)^2   \Delta_{\rm st} ( \kvec) 
\end{align}
where we have absorbed factors of $f$ into the time dependent coefficients.  Notice that in the stochastic terms, higher powers in $\mu$ are accompanied by higher powers in $k$, so we have only included the lowest order here, and $c_{\rm st} \Delta_{\rm st} = \half ( c_{\rm st,1}\Delta^{ij}_{\rm st , 1} ( \kvec )  + c_{\rm st , 2}\Delta^{ij}_{\rm st , 2})\zhat_i \zhat_j $.  This leads to a contribution to the power spectrum of the form
\begin{align} \nonumber
P^r_{k^4, \rm c.t.}  = & - (2\pi)^2 D^2 ( 1 + f \mu^2) \left( \tilde c_1^2  \mu^2 + \tilde c_2^2  \mu^4 +  \tilde c_3^2  \mu^6 \right)  \left( \frac{k}{\knlv} \right)^4 P_{11}^r ( k )  \\
& +\frac{(2 \pi)^2}{4} D^2  \left( \bar c_1^2 \mu^2 + \bar c_2^2 \mu^4 \right)^2  \left( \frac{k}{\knlv} \right)^4 P_{11}^r ( k ) +c^2_{\rm st}  \mu^4 \left( \frac{ k }{\knlv} \right)^4 \frac{ (2 \pi)^2 }{k_{\rm NL}^{r \, 3}}  \ ,
\end{align}
where the first term on the second line comes from squaring the piece already present in \eqn{deltarr}. We have redefined $c_{\rm st}$ to absorb the parameters $C^{(i)}_{ab}$, and we have redefined all of the $c^2$ constants by $c^2 \rightarrow (2 \pi)^2 c^2$.  Indeed, the leading stochastic piece in the power spectrum is proportional to $(k / \knlv)^4$. 

In Figure~\ref{k4plots}, we take $\knlv = \knl / 2.5$ as an example, and present the results at $z=0.56$.  The parameters that we use in Figure~\ref{k4plots} are  
\begin{align}\label{eq:param1}
\bar c_1^2 & = -0.54 \,  (\knl / \unitsk)^2  = - 3.4  \, ( \knlv  / \unitsk)^2 \\ \nonumber
\bar c_2^2 & = 0.076 \, ( \knl  / \unitsk)^2 = 0.48 \, ( \knlv  / \unitsk)^2  \\\nonumber
\tilde c_1^2 & =0.13 \, ( \knl  / \unitsk)^4 =   5.1 \, ( \knlv  / \unitsk)^4  \\ \nonumber
\tilde c_2^2 &= -0.035 \, ( \knl  / \unitsk)^4 =  1.4  \, ( \knlv  / \unitsk)^4   \\\nonumber
\tilde c_3^2 & =0.026 \, ( \knl  / \unitsk)^4 =  1.0   \, ( \knlv  / \unitsk)^4      \ ,
\end{align} 
 which are more comparable to the dark-matter speed of sound $c_s^2 = 0.31  \, ( \knl  / \unitsk)^2$ after we assumed that $\knlv = \knl / 2.5$.  With these parameters, the theory fits until $k~\approx~0.18 \unitsk$. The parameters that we use in Figure~\ref{k4plots}, including the stochastic term, are  
  \begin{align}\label{eq:param2}
\bar c_1^2 & = -0.54 \,  (\knl / \unitsk)^2  = - 3.4  \, ( \knlv  / \unitsk)^2 \\  \nonumber
\bar c_2^2 & = 0.072 \, ( \knl  / \unitsk)^2 = 0.45 \, ( \knlv  / \unitsk)^2  \\\nonumber
\tilde c_1^2 & =0.14 \, ( \knl  / \unitsk)^4 =   5.5 \, ( \knlv  / \unitsk)^4  \\ \nonumber
\tilde c_2^2 &= -0.041 \, ( \knl  / \unitsk)^4 =  1.6  \, ( \knlv  / \unitsk)^4   \\\nonumber
\tilde c_3^2 & =0.053 \, ( \knl  / \unitsk)^4 =  2.1   \, ( \knlv  / \unitsk)^4 \\\nonumber
c_{\rm st}^2 & = 5.0 \, ( \knl  / \unitsk)^4 =  195   \, ( \knlv  / \unitsk)^4      \ . 
\end{align}  
In order for the stochastic term to have an effect, we needed $c_{\rm st}^2 = 5.0 \, ( \knl  / \unitsk)^4$, which is much larger than the other coefficients.  Even with the large parameter, the fit is not significantly improved.  Thus, unless there is a reason for the coefficient to be this large, the stochastic piece can be neglected at this order.

\begin{figure}[htb!] 
\begin{center}
\includegraphics[width=8cm]{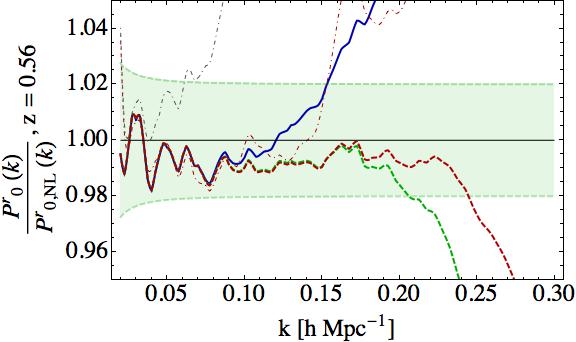} \includegraphics[width=8cm]{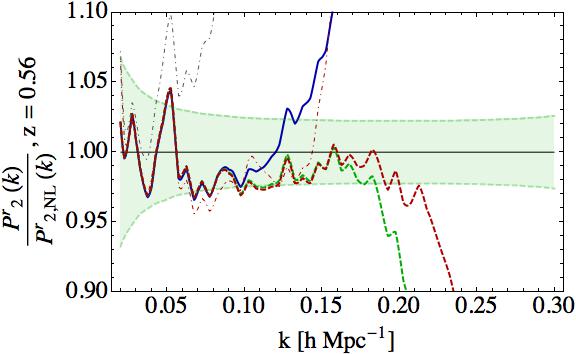} \\
\includegraphics[width=8cm]{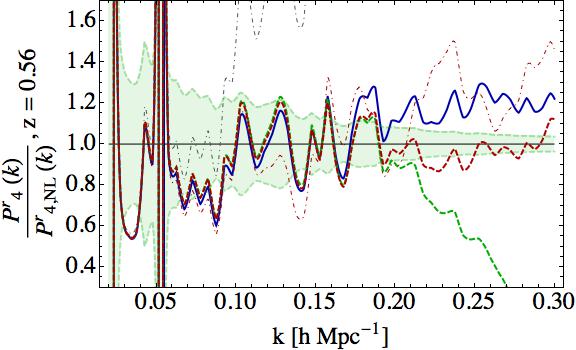} \includegraphics[width=8cm]{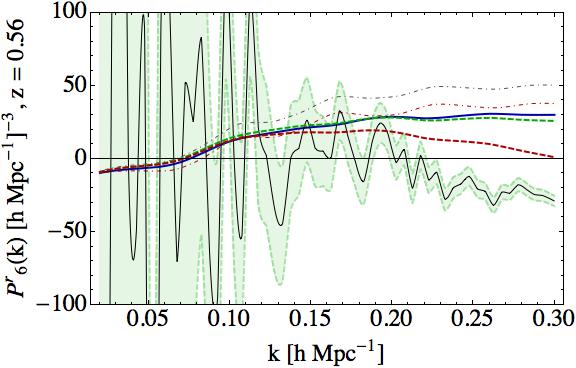}
\caption{Here, we show two possible improvements in the small $\knlv$ limit at $z=0.56$.  The dark blue curve is the one-loop IR-resummed EFT prediction with  $\bar c_1^2=-0.56\,(\knl / \unitsk)^2$ and $\bar c_2^2=0.13\,( \knl  / \unitsk) ^2$, the same as in Figure~\ref{pls} but we have set $\knlv=\knl / 2.5$.  The dashed green curve includes the $( k / \knlv)^4 P_{11}$ counter terms with no stochastic piece and the remaining parameters given by \eqn{eq:param1}.  The dashed red curve includes the stochastic piece and has parameters $\bar c_1^2=-0.54\,(\knl / \unitsk)^2$, $\bar c_2^2=0.072\,( \knl  / \unitsk)^2$ and the remaining parameters given by~\eqn{eq:param2}.  Thus, including these higher order terms can extend the fits until $k \approx 0.18 \unitsk$ without the stochastic term, and to $k \approx 0.2 \unitsk$ with the stochastic term.  The dot-dashed red curve, the gray dot-dashed curve,  and the green region are the same as in Figure~\ref{pls}.  }  \label{k4plots}
\end{center}
\end{figure}

\section*{Acknowledgments}

We would like to thank Matias Zaldarriaga for discussions. C. C. acknowledges support as a MultiDark Fellow. L.S. is partially supported by DOE Early Career Award DE-FG02-12ER41854.  F.P., C.Z., and C.C. acknowledge support from the Spanish MICINNs Consolider-Ingenio 2010 Programme under grant MultiDark CSD2009-00064, MINECO Centro de Excelencia Severo
 Ochoa Programme under grant SEV-2012-0249, and grant AYA2014-60641-C2-1-P.  FP wishes to thank KIPAC for the hospitality during the development of this work.
The BigMultidark simulations have been performed on the SuperMUC supercomputer at the Leibniz-Rechenzentrum (LRZ) in Munich, using the computing resources awarded to the PRACE project number 2012060963.

\section*{Appendices}

\begin{appendix}

\section{Conventions and Legendre Polynomial Identities}
 In this paper, we use the standard conventions for the power spectrum in redshift space with line of sight $\hat z$
\begin{align} \label{legendre}
P^r ( k , \kdotz) & = \sum_{\ell  } \mathcal{P}_\ell ( \kdotz ) P_\ell^r ( k) \\
P^r_\ell ( k ) & = \frac{2 \ell + 1}{2} \int_{-1}^1 d \mu \, \mathcal{P}_\ell (\mu) P^r ( k , \mu)  \\
\xi^r ( q , \qdotz) & = \int \momspmeas{k} e^{i \vec{k} \cdot \vec{q} } P^r ( k , \kdotz) = \sum_{\ell} \mathcal{P}_\ell ( \qdotz ) \xi^r_\ell ( q) \\
\xi^r_\ell ( q ) & = i^\ell \int dk \, \frac{ k^2}{2 \pi^2} P_\ell^r ( k ) j_\ell ( k q) \\
P_\ell^r ( k ) & = 4 \pi ( - i )^\ell \int dq \, q^2 \xi^r_\ell (q ) j_\ell ( k q ) \ ,
\end{align}
where $P^r$ is the power spectrum in redshift space, $\xi^r$ is the correlation function in redshift space, and $\cp_\ell$ is the $\ell$th Legendre polynomial.  The first few Legendre polynomials are $\cp_0 (\mu)=1$, $\cp_1(\mu) =\mu$, and $\cp_2 ( \mu ) = (3 \mu^2 - 1) / 2$.  A few useful identities are
\begin{align} \label{planewave}
   \sum_{\ell}^\infty i^\ell ( 2 \ell + 1) j_\ell ( k q ) \mathcal{P}_\ell (\kdotq) & = e^{i k q (\kdotq) } \\
\int_0^\infty dr \, r^2 \, j_\ell( k r ) j_\ell ( k' r ) & = \frac{\pi}{ 2 k^2} \delta( k - k') \\
\int d^2 \hat q  \, \mathcal{P}_{\ell' } ( \hat k \cdot \hat q ) \mathcal{P}_{\ell} ( \hat q \cdot \hat z ) & = \frac{4\pi}{2\ell +1} \delta_{\ell \ell'} \mathcal{P}_\ell ( \hat k \cdot \hat z) \label{legpoly1}
\end{align}
where $j_\ell$ is the spherical Bessel function of type $\ell$.

\section{IR-resummation in Real Space} \label{ircompare}
In this Appendix, we run some consistency checks on the new method of IR-resummation for the real space dark-matter power spectrum.  In terms of the notation used in this paper, this is given by 

\begin{align} \nonumber
P ( k ) \Big|_N & = \sum_{j = 0}^N  \Bigg( \int \frac{d k' \, k'^2}{2 \pi^2} \Delta M_{||_{N-j}} ( k , k')_{00} P ( k' )_j  \\
&\hspace{.5in} +  \frac{1}{2} e^{-\frac{k^2}{2} X_1^\infty }  \left[ \kappa_{||_{N-j}}^{(n_Y)} (k , \infty) \right]_{a00}   I^a_{00} ( k , \infty) P ( k ) _j \Bigg) \Bigg|_{f=0} 
\end{align}
In the real-space case, $ \bar K_0  ( \vec{k} , \qvec  )  \cdot K_0^{-1} ( \vec{  k} , \qvec ) \Big| \Big|_{N-j} $ does not depend on $\mu_k$ or $\mu_q$, so $\left[ \kappa_{||_{N-j}}^{(n_Y)} (k , q) \right]_{abc}$ is zero unless $a=0$ and $b=0$.  Furthermore, 
\be
 \left[ \kappa_{||_{N-j}}^{(n_Y)} (k , \infty) \right]_{000} \Bigg|_{f=0}  = e^{\frac{k^2}{2} X_1^\infty } \Big|_{N-j}
\ee
is just the expansion of $e^{\frac{k^2}{2} X_1^\infty }$ to order $N-j$.  This leads to 
\begin{align} 
P ( k ) \Big|_N & = \sum_{j = 0}^N  \Bigg(   \int \frac{d k' \, k'^2}{2 \pi^2} \Delta M_{||_{N-j}} ( k , k')_{00} P ( k' )_j  + e^{-\frac{k^2}{2} X_1^\infty }   \left[ \kappa_{||_{N-j}}^{(n_Y)} (k , \infty) \right]_{000}   P ( k ) _j   \Bigg) \Bigg|_{f=0}
\end{align}
where 
\begin{align} \nonumber
\Delta M_{||_{N-j}} ( k , k')_{00} &= 4 \pi \int d q \, q^2 j_0 ( k' q ) \Bigg(   e^{-\frac{k^2}{2} \left( X_1(q) + \alpha Y_1 ( q) \right) }  \sum_c i^c j_0^{(c)} ( k q) \left[ \kappa_{||_{N-j}}^{(n_Y)} (k , q) \right]_{00c}    \\
&    \hspace{2in}  - e^{-\frac{k^2}{2} X_1^\infty }  \left[ \kappa_{||_{N-j}}^{(n_Y)} (k , \infty) \right]_{000}   j_0 ( kq)    \Bigg) \Bigg|_{f=0} \ . 
\end{align}
Thus, after the expansion in terms of $k^2 Y_1$, the resummation has been simplified by performing the integral over $d^2 \hat q$ in terms of simple analytical functions.  The accuracy of this approach depends only on $n_Y$, and can be made arbitrarily accurate by expanding to higher orders.  Before the integral over $d q$, we only need to perform the fast analytic manipulations of computing $j^{(c)}_0$ and extracting the coefficients $\left[ \kappa_{||_{N-j}}^{(n_Y)} (k , q) \right]_{00c}$.  In Figure~\ref{comparey3}, we show the the result for $n_Y =1,2,3$, and see that the result changes only slightly from $n_Y = 1$ to $n_Y = 2$, and changes by an imperceptible amount from $n_Y = 2$ to $n_Y = 3$.  

\begin{figure}[htb!] 
\begin{center}
\includegraphics[width=11cm]{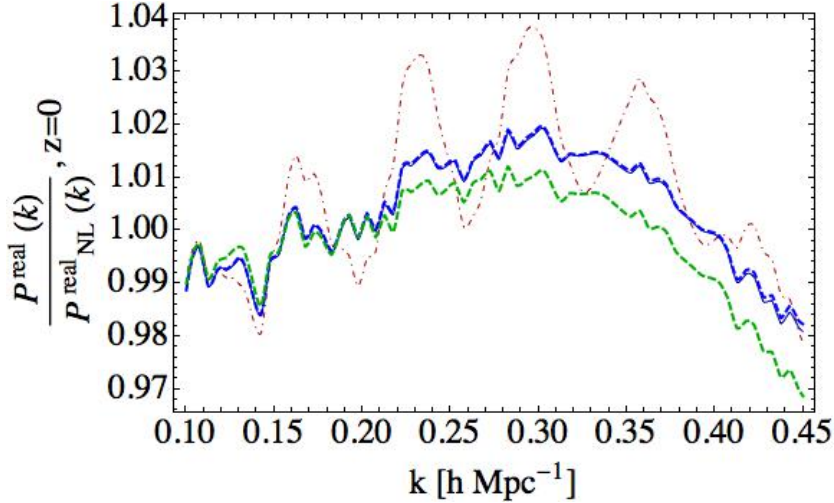} 
\caption{This plot shows the convergence of our approximation scheme in the $k^2 Y_1$ expansion, and compares it to the method of resummation in \cite{Senatore:2014via, Foreman:2015uva}.  The thin solid blue curve is the method presented in this paper at one loop with $n_Y =1$ (it is barely distinguishable from the thick dashed blue curve), the thick dashed blue curve is the method in this paper with $n_Y=2$ and $n_Y=3$ (they are indistinguishable on this plot), the thick green dashed curve is the method from \cite{Senatore:2014via, Foreman:2015uva}, and the dot-dashed red curve is the non-resummed, one-loop EFT power spectrum.  The value of $c_s^2$ has been chosen to emphasize the resummation, not to correspond to the physical theory.  }  \label{comparey3}
\end{center}
\end{figure}

In Figure~\ref{comparey3}, we also plot the result of the method of resummation in \cite{Senatore:2014via, Foreman:2015uva}, where we see that the two are essentially identical except for an offset at high $k$.  This offset is due to the fact that, at one loop, both of the resummation techniques are only accurate to order $\epsilon_{\delta <}$, and can differ by an amount $\epsilon_{\delta < }^2$.   The difference is recovered order by order in the loop expansion.

\section{Results at $z=0$} \label{z0}
In Figure~\ref{dmz0} we present our fit to the BigMultiDark dark-matter power spectrum~\cite{Klypin:2014kpa} at $z=0$, and in Figure~\ref{lsz0} we show our results in redshift space.  We find that the theory fits until about $k\approx 0.12 \unitsk$ if we only include the quadratic counter terms, and  it fits until about $k \approx 0.15 \unitsk$ if we include the $( k / \knlv )^4$ counter terms.  We also find that the stochastic terms have a very small effect on the final result, even if the coefficient is large.  In Figure~\ref{lsz0}, we show the prediction with only quadratic counter terms, using the values 
\begin{align}
c_s^2  &  = 0.20  \, (\knl / \unitsk)^2 \\ \nonumber
\bar c_1^2 & = 0.33 \,  (\knl / \unitsk)^2 = 2.1 \,  (\knlv / \unitsk)^2 \\\nonumber
\bar c_2^2 & = 0.080 \, ( \knl  / \unitsk) ^2 = 0.50  \,  (\knlv / \unitsk)^2
\end{align} 
where we have assumed $\knlv = \knl / 2.5$ as in Section~\ref{smallkv}.  When including the leading $(k / \knlv)^4 P_{11}$ counter terms, in Figure~\ref{lsz0} we use
\begin{align}\label{eq:parameter3}
 \bar c_1^2 & = 0.30 \,  (\knl / \unitsk)^2   \\ \nonumber
 \bar c_2^2 & = 0.074 \, ( \knl  / \unitsk)^2  = 0.46 \,  (\knlv / \unitsk)^2    \\\nonumber
 \tilde c_1^2 & =0.21 \, ( \knl  / \unitsk)^4   =   8.2 \,  (\knlv / \unitsk)^4 \\\nonumber
 \tilde c_2^2 & = -0.0023 \, ( \knl  / \unitsk)^4 =   0.090\,  (\knlv / \unitsk)^4 \\\nonumber
 \tilde c_3^2 & = - 0.026 \, ( \knl  / \unitsk)^4   =   1.0 \,  (\knlv / \unitsk)^4\ . 
\end{align} 
As at $z=0.56$, we find that in order for the stochastic piece to have any effect, we need $c_{\rm st}^2 = 4.00 \, ( \knl  / \unitsk)^4$ (with the other coupling constants roughly the same, see Figure~\ref{dmz0}), which is much larger than the coupling constants above.  Even when the coefficient is this large, the effect is small, so we conclude that the stochastic piece does not improve our fits at this order.  For completeness, in this case the coefficients are given by
 \begin{align}\label{eq:parameter4}
 \bar c_1^2 & = 0.30 \,  (\knl / \unitsk)^2   \\ \nonumber
 \bar c_2^2 & = 0.063 \, ( \knl  / \unitsk)^2  = 0.39 \,  (\knlv / \unitsk)^2    \\\nonumber
\tilde c_1^2 & =0.23\,( \knl  / \unitsk)^4  =   9.0 \,  (\knlv / \unitsk)^4 \\\nonumber
\tilde c_2^2 & =0.0068\,( \knl  / \unitsk)^4 =   0.27\,  (\knlv / \unitsk)^4 \\\nonumber
\tilde c_3^2 & =-0.018\,( \knl  / \unitsk)^4  =   0.70\,  (\knlv / \unitsk)^4  \\ \nonumber 
c_{\rm st}^2\, & = 4.00\,( \knl  / \unitsk)^4   =    156 \,  (\knlv / \unitsk)^4   \ .  
\end{align}

\begin{figure}[htb!] 
\begin{center}
\includegraphics[width=10cm]{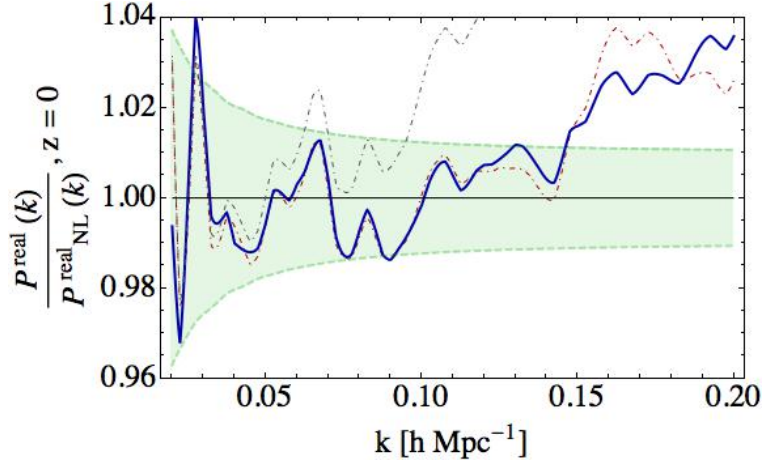} 
\caption{ Here, we compare the EFT prediction with $c_s^2  = 0.20  \, (\knl / \unitsk)^2$ to the BigMultiDark dark-matter power spectrum~\cite{Klypin:2014kpa} at $z=0$.  The dark blue curve is the one-loop IR-resummed EFT prediction, the red dot-dashed curve is the non-resummed EFT prediction, the gray dot-dashed curve is one-loop SPT, and the green region is the error on the data, where we have included a $1\%$ systematic error.  }  \label{dmz0}
\end{center}
\end{figure}

\begin{figure}[htb!] 
\begin{center}
\includegraphics[width=8cm]{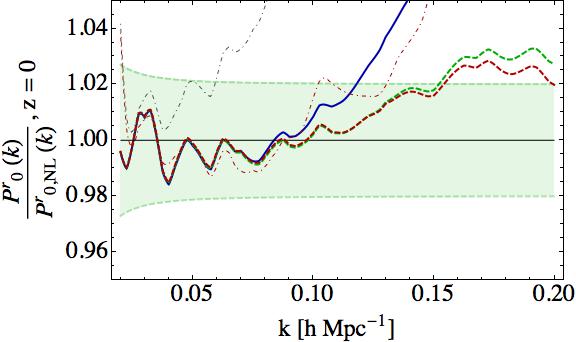} \includegraphics[width=8cm]{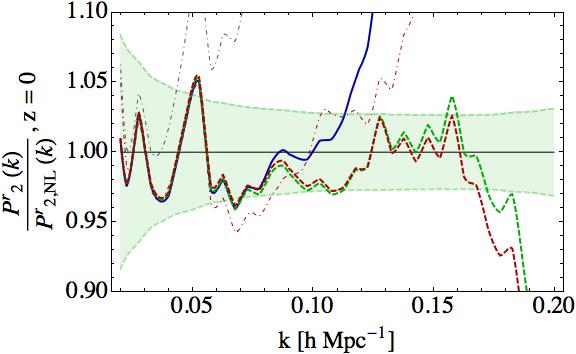} \\
\includegraphics[width=8cm]{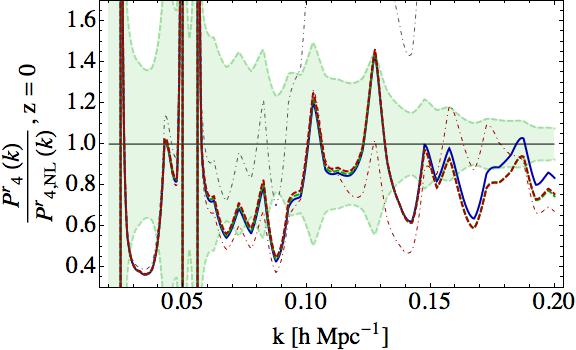} \includegraphics[width=8cm]{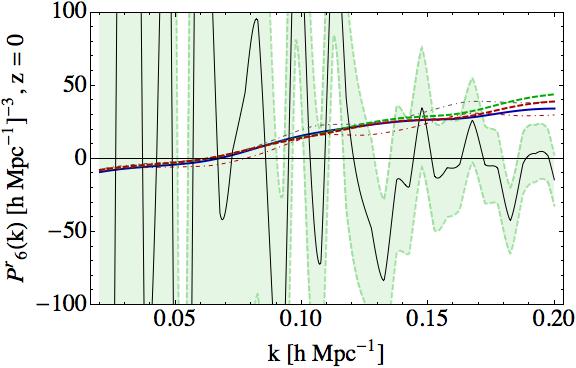}
\caption{ 
 Here, we summarize our results at $z=0$.  The dark blue curve is the one-loop IR-resummed EFT prediction with  $\bar c_1^2=0.33\,(\knl / \unitsk)^2$ and $\bar c_2^2=0.080\,( \knl  / \unitsk)^2$.  The dashed green curve includes the $( k / \knlv)^4 P_{11}$ counter terms with no stochastic piece and parameters $\bar c_1^2=0.30\,(\knl / \unitsk)^2$, $\bar c_2^2=0.074\,( \knl  / \unitsk)^2$, and the rest of the parameters given by~\eqn{eq:parameter3}. The dashed red curve includes the stochastic piece and has parameters $\bar c_1^2=0.30\,(\knl / \unitsk)^2$, $\bar c_2^2=0.063\,( \knl  / \unitsk)^2$, and the rest of the parameters given by~\eqn{eq:parameter4}. Including only $k^2$ counter terms, the EFT fits until about $k \approx 0.12 \unitsk$.  Including the $k^4$ terms can extend the fits until $k \approx 0.15 \unitsk$ without the stochastic term; the stochastic term makes only a small difference.  The dot-dashed red curve, the gray dot-dashed curve,  and the green region are the same as in Figure~\ref{pls}.   }  \label{lsz0}
\end{center}
\end{figure}

\section{Freeing the Time Dependence \label{freetime}}

In redshift space, the time dependence of $c_s^2$ is important, even at a single redshift, because $\dot c_s^2$ enters in \eqn{prred}.   In a scaling universe with slope $n$, the time dependence of $c_s^2$ is determined by the scaling symmetry to be $c_s^2 ( t ) \propto D^{ 4 / ( 3 +n) }$.  However, in the real universe, this time dependence is only approximate.  Thus, we don't know exactly how  $\dot c_s^2$ is related to $c_s^2$ in \eqn{prred}. In general it should be included as an additional parameter of the effective field theory.  To parametrize the time dependence we make the replacement 
\be
\frac{d\, c_s^2}{d\log a} \rightarrow \frac{d\, c_s^2}{d\log a} ( 1 + \gamma)
\ee
and expect $\gamma \lesssim \mathcal{O}(1)$ because of the approximate scaling symmetry.  We find that including this does not significantly change any of the fits.  Figures~\ref{timez0}~and~\ref{timez56} show the results at $z=0$ and $z=0.56$ respectively.

\begin{figure}[htb!] 
\begin{center}
\includegraphics[width=8cm]{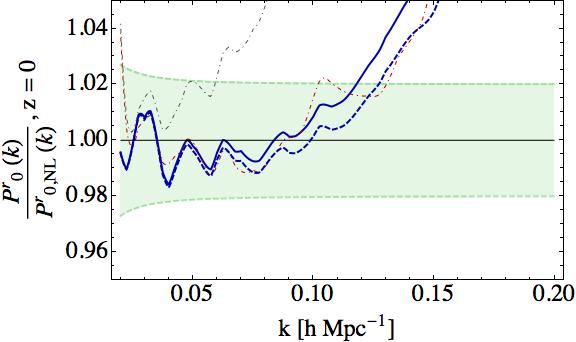} \includegraphics[width=8cm]{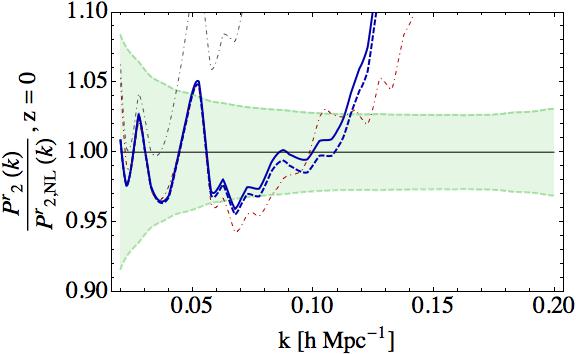} \\
\includegraphics[width=8cm]{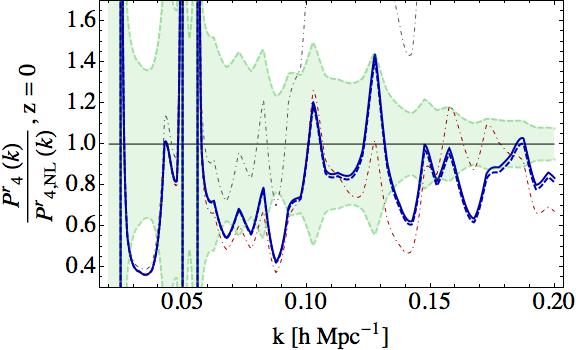} \includegraphics[width=8cm]{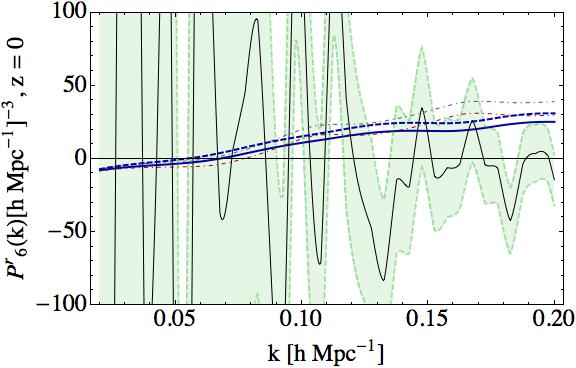} \\
\caption{  We present the results of freeing the time dependence of $c_s^2$ at $z=0$.  The thick blue curve is best fit for $z=0$ from Figure~\ref{lsz0}: $c_s^2=0.20\,(\knl / \unitsk)^2$, $\bar c_1^2=2.1\,(\knlv / \unitsk)^2$,  $\bar c_2^2~=0.50\,( \knlv  / \unitsk) ^2$.  The dashed blue curve is the new fit with $c_s^2=0.20\,( \knl  / \unitsk)^2$, $\bar c_1^2=2.4\,( \knlv  / \unitsk)^2$, $\bar c_2^2=0.13\,( \knlv  / \unitsk)^2$, and $\gamma=0.5$.   }  \label{timez0}
\end{center}
\end{figure}

\begin{figure}[htb!] 
\begin{center}
\includegraphics[width=8cm]{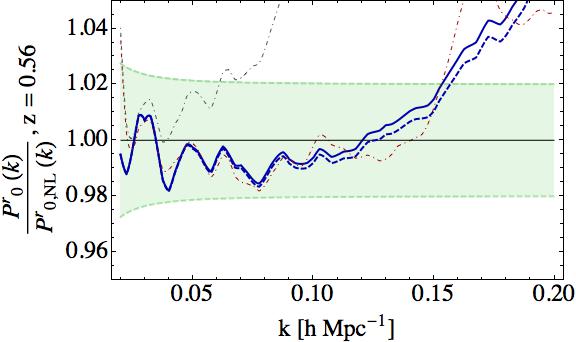} \includegraphics[width=8cm]{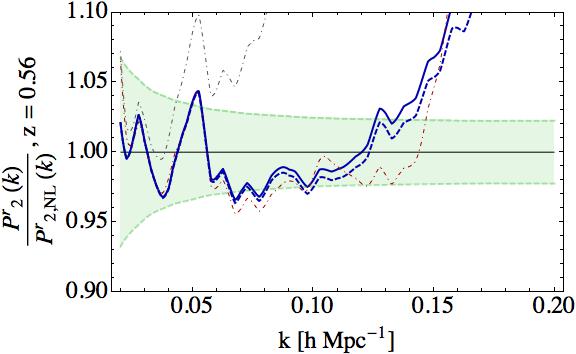} \\
\includegraphics[width=8cm]{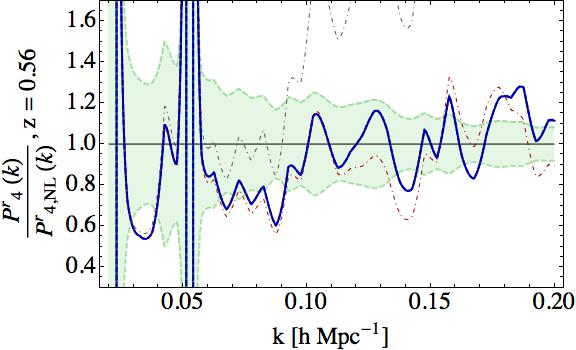} \includegraphics[width=8cm]{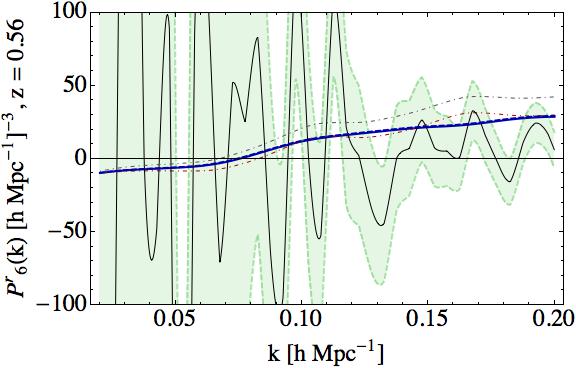} \\
\caption{   We present the results of freeing the time dependence of $c_s^2$ at $z=0.56$.  The thick blue curve is best fit for $z=0.56$ from Figure~\ref{pls}:  $ c_s^2=0.31\,( \knl  / \unitsk) ^2$, $\bar c_1^2=-3.5\,(\knlv / \unitsk)^2$, and $\bar c_2^2=0.81\,( \knlv  / \unitsk) ^2$.  The dashed blue curve is the new fit with $c_s^2=0.31\,( \knl  / \unitsk)^2$, $\bar c_1^2=-3.9\,( \knlv  / \unitsk)^2$ , $\bar c_2^2=0.76\,( \knlv  / \unitsk)^2$, and $\gamma=0.09$.   }  \label{timez56}
\end{center}
\end{figure}

\end{appendix}

 \begingroup\raggedright\endgroup

 \end{document}